\documentclass[10pt,journal]{IEEEtran}
\usepackage{cite}
\usepackage{amsmath,amssymb,amsfonts}
\usepackage{algorithmic}
\usepackage{graphicx}
\usepackage{textcomp}
\usepackage{ulem}
\usepackage[hyphens]{url}
\usepackage[braket, qm]{qcircuit}
\usepackage{graphicx}
\usepackage{caption}
\def\BibTeX{{\rm B\kern-.05em{\sc i\kern-.025em b}\kern-.08em
    T\kern-.1667em\lower.7ex\hbox{E}\kern-.125emX}}
\newcommand{\cmark}{\ding{51}}%
\newcommand{\xmark}{\ding{55}}%
\usepackage{color,xcolor}
\usepackage{xspace}
\usepackage{multirow}
\usepackage{booktabs}
\usepackage{lipsum}
\usepackage{soulutf8}
\usepackage{array}
\usepackage[caption=false,font=normalsize,labelfont=sf,textfont=sf]{subfig}
\usepackage{textcomp}
\usepackage{stfloats}
\usepackage{url}
\usepackage{verbatim}
\usepackage{graphicx}
\usepackage{cite}
\usepackage{colortbl}
\usepackage{float}
\usepackage{placeins}

\usepackage{enumitem}
\usepackage{braket}

\usepackage{amsmath}
\usepackage{amsfonts}
\usepackage{url}

\usepackage{afterpage}
\usepackage{ulem}
\usepackage[ruled,vlined]{algorithm2e}
\usepackage{setspace}

\usepackage{pifont}

\usepackage{xcolor}
\usepackage{soul}
\usepackage{hyperref}

\begin{document}
\title{NAPA: Intermediate-level Variational \underline{Na}tive-\underline{p}ulse \underline{A}nsatz for Variational Quantum Algorithms}
\author{
    \IEEEauthorblockN{Zhiding Liang\textsuperscript{1},
    Jinglei Cheng\textsuperscript{1},
    Hang Ren,
    Hanrui Wang,
    Fei Hua,
    Zhixin Song,
    Yongshan Ding,
    Frederic T. Chong,
    Song Han,
    Xuehai Qian,
    Yiyu Shi\\
\textsuperscript{1}These authors contributed to the work equally and should be regarded as co-first authors. \\
}
\vspace{-18pt}
    \thanks{Z. Liang and Y. Shi are with the Department of Computer Science and
Engineering, University of Notre Dame, Notre Dame, IN, 46556 USA.

J. Cheng and X. Qian are with the Department of Computer Science, Purdue University, West Lafayette, IN 47907 USA.

H. Ren is with the Department of Chemistry, University of California, Berkeley
Berkeley, CA 94720 USA.

H. Wang and S. Han are with the Department of EECS, Massachusetts Institute of Technology,  Cambridge, MA 02139 USA.
F. Hua is with the Department of Computer Science, Rutgers University, New Brunswick, NJ 08901 USA.

Z. Song is with the Department of Physics, Georgia Institute of Technology, Atlanta, GA 30332 USA.

Y. Ding is with the Department of Computer Science, Yale University, New Haven, CT 06511 USA.

F. Chong is with the Department of Computer Science, University of Chicago, Chicago, Illinois 60637 USA.

Address comments to Zhiding Liang (email: zliang5@nd.edu).
}% <-this % stops a space
}
%%%%%%%%%%%%%%%%%%%%%%%%%%%%%%%%%%%%

\maketitle
\thispagestyle{plain}
\pagestyle{plain}

\newcommand{\jlc}[1]{\textcolor{violet}{[jinleic: #1]}}
\newcommand{\noteYD}[1]{\textcolor{blue}{[YD: #1]}}
\newcommand{\name}{\texttt{NAPA}}
\newcommand{\red}[1]{\textcolor{black}{#1}}
\newcommand{\revise}[1]{\textcolor{black}{#1}}

\newcommand{\rebuttal}[1]{\textcolor{black}{#1}}

%%%%%% -- PAPER CONTENT STARTS-- %%%%%%%%
\begin{abstract}
Variational quantum algorithms (VQAs) have demonstrated great potentials in the \revise{Noisy Intermediate Scale Quantum} (NISQ) era.
In the workflow of VQA, the parameters of ansatz are iteratively updated to approximate the desired quantum states.
We have seen various efforts to draft better ansatz with less gates. 
Some works consider the physical meaning of the underlying circuits, while others adopt the ideas of neural architecture search (NAS) for ansatz generator.
However, these designs do not exploit the full advantages of VQAs. 
Because most techniques target gate ansatz, and the parameters are usually rotation angles of the gates.
In quantum computers, the gate ansatz will eventually be transformed into control signals such as microwave pulses on \revise{superconducting qubits}.
These control pulses need elaborate calibrations to minimize the errors such as over-rotation and under-rotation.
In the case of VQAs, this procedure will introduce redundancy, but the variational properties of VQAs can naturally handle problems of over-rotation and under-rotation by updating the amplitude and frequency parameters.
Therefore, we propose \name, a native-pulse ansatz generator framework for VQAs.
We generate native-pulse ansatz with trainable parameters for amplitudes and frequencies.
In our proposed \name, we are tuning parametric pulses, which are natively supported on NISQ computers.
\rebuttal{Given the limited availability of gradient-based optimizers for pulse-level quantum programs, we choose to deploy non-gradient optimizers in our framework.}
To constrain the number of parameters sent to the optimizer, we adopt a progressive way to generate our native-pulse ansatz.
Experiments are conducted on both simulators and quantum devices \revise{for Variational Quantum Eigensolver (VQE) tasks} to envaluate our methods.
When adopted on NISQ machines, \name \space obtained improved the performance with decreased latency by an average of 86\%.
\name \space is able to achieve 96.482\% and 99.336\% accuracy for VQE tasks on $H2$ and $HeH+$ respectively.
An average accuracy of 97.27\% is achieved for medium-size quantum chemistry tasks on $CO_2$, $H_2O$, and $NaH$. 
\name~also demonstrates advantages on quantum optimization tasks even with considerable noises in NISQ machines.
\end{abstract}

\section{Introduction}
\label{sec1}
%DEVELOPMENTS OF TODAY'S QUANTUM COMPUTING
Operating on the principles of quantum mechanics, quantum computers have the potential to solve problems that are intractable on classical computers~\cite{aharonov2008fault}.
As hardware technologies and quantum algorithms advance rapidly, today's quantum computers begin to demonstrate their potentials in solving problems of non-trivial size in areas such as quantum chemistry~\cite{axelrod2022excited}.
In 2019, Google claimed to have achieved quantum supremacy with the task of random circuit sampling on a 53-qubit quantum computer~\cite{arute2019quantum}.  
\rebuttal{IBM introduced the Condor, a 1,121 superconducting qubit quantum processor at the end of 2023, together with a roadmap for modular quantum processors with more than 10,000 qubits in the coming years~\cite{ibm1000qubits}. Recent studies have demonstrated promising approaches for constructing logical quantum processors using error correction techniques on reconfigurable atom arrays~\cite{bluvstein2023logical}.} 
%Currently available quantum computers have demonstrated capabilities to solve problems of non-trivial size in areas such as quantum chemistry~\cite{sauceda2022bigdml, axelrod2022excited}.
In the current NISQ era, however, emerging quantum devices are still prone to errors and sensitive to decoherence~\cite{ding2020quantum}. 

% NISQ ERA AND DEVICES
In superconducting quantum computers, qubits are sparse and not perfectly isolated from \revise{their} environments. These quantum devices have not yet met the hardware requirements for error correction methods such as surface codes~\cite{das2022afs} due to the limited number of qubits and low gate fidelity.
Without quantum error correction, current quantum computers can only handle small-scale circuits before running into irreversible errors, making practical-size algorithms infeasible.
However, with elaborately designed noise-resilient algorithms, we can still expect to achieve quantum advantages in areas such as quantum chemistry~\cite{tomesh2021optimized,cao2019quantum} much sooner than other applications like database search~\cite{grover1996fast} and integer factorization~\cite{shor1999polynomial}.

%VQA(VQE QAOA QNN)
Variational quantum algorithms~\cite{cerezo2021variational, lubasch2020variational} have shown great noise resilience, and are considered as hybrid algorithms, where some parts are performed on a quantum device and others on a classical computer.
Variational quantum eigensolver (VQE)~\cite{kandala2017hardware, wang2019accelerated} is one of the most promising candidates in the variational computing paradigm. 
With VQE, we are able to estimate the ground state energy of a targeting quantum system by iteratively updating a parametrized quantum circuit.
Quantum approximate optimization algorithm (QAOA)~\cite{guerreschi2019qaoa} and quantum neural network (QNN)~\cite{abbas2021power} are also members of VQAs.
\revise{QAOA attempts to solve combinatorial optimization problems, including maxcut problems, while QNN has exhibited exceptional capabilities in representing complex data~\cite{ herrmann2022realizing}.}
These algorithms are among the most promising examples of NISQ algorithms since the number of required quantum gates remains moderate.

%COMPILATION WORKFLOW
\noindent\textbf{Exposing Native Pulse-Level Controls.}\
The majority of existing quantum computers do not provide access for analog controls of qubits.
Consequently, nearly all compilers implement a gate-based workflow~\cite{tomesh2022supermarq}, in which quantum algorithms are synthesized, compiled on classical computers, and finally executed on quantum computers.
In the first place, quantum circuits are generated or synthesized to implement certain functions of quantum algorithms.
These gate circuits are usually not compatible with the underlying topology of quantum hardware.
Therefore, SWAP gates need to be inserted to make quantum circuits executable on quantum computers.
Then these circuits are decomposed into single-qubit and two-qubit gates that are natively supported by quantum computers.
Finally, the circuits are dispatched to quantum backends, where they are ``translated'' into control signals on physical qubits such as transmons~\cite{place2021new}, trapped ions~\cite{bruzewicz2019trapped} and photons~\cite{kok2007linear}.
For superconducting quantum computers, the control signals are microwave pulses~\cite{alexander2020qiskit}.

%\begin{figure}[t]
%\centering
%\includegraphics[width=\linewidth]{figures/ansatz.pdf}
%\caption{A traditional procedure for gate-based quantum variational algorithms. During each iteration, gradient information is used to update the ansatz's parameters. Gradients are ``calculated'' using finite difference techniques or \red{parameter-shift} rules.}
% \vspace{-1mm}
%\label{fig_ansatz}
%\end{figure}

\begin{table*}[t]
    \renewcommand*{\arraystretch}{1.4}
    \setlength{\tabcolsep}{14pt}
    \footnotesize
    \begin{center}
    \begin{tabular}{|l|cc|cc|ccc|}
        \hline
        \multicolumn{1}{|c|}{Method} & \multicolumn{2}{c|}{Robustness} & \multicolumn{2}{c|}{Parameters of Control} & \multicolumn{3}{c|}{Applications}\\
         & Noise-aware & System model & Amp. & Freq.  & VQE & QAOA & QML \\
        \hline
        Gate-level (conventional)  & \cmark & \cmark & \xmark & \xmark &\cmark & \cmark&\cmark  \\
        Pulse-level attempts~\cite{meitei2021gate, choquette2021quantum}  & \xmark & \xmark &\cmark & \cmark    &\cmark & \xmark &\xmark  \\
        \hline
        \textbf{\name~(Proposed)}  & \cmark & \cmark & \cmark & \cmark  &\cmark & \cmark&\cmark \\
        \hline
    \end{tabular}
        \caption{Comparison between gate-level approaches, pulse-level approaches, and the proposed \name}
    %\HW{add [21] comparison if possible; add differential pulse simulator?}
    % \HW{put all x marks to the front}
    \vspace{-15pt}
    \label{table_comp}
    \end{center}
\end{table*}

%PROS and CONS PULSES
\red{There is more flexibility if gate circuits are decomposed and controlled at pulse level, since we are dealing with a more fine-grained abstraction layer.}
The challenge now is to figure out how to effectively generate pulses for quantum algorithms.
Existent techniques such as quantum optimal control (QOC)~\cite{ meitei2021gate, choquette2021quantum} can be adopted to generate control pulses. 
QOC devises and implements the shapes of external controls on qubits to accomplish given tasks.
As indicated in \cite{shi2019optimized}, QOC can handle quantum circuits of moderate size, but the scalability might be the issue.
Despite various efforts~\cite{cheng2020accqoc} to optimize QOC and reduce its overhead, it is still computationally expensive.
% Also, we need to consider the high-noise feature of NISQ devices, on which it is preferred to have a fixed set of allowed operations or gates. 
% This small group of operations or gates can be carefully calibrated to reach the maximum fidelity~\cite{alexander2020qiskit}. 
% The calibrations are conducted regularly to enforce high accuracy.
Besides, we need to consider the high-noise feature of NISQ devices, on which it is preferred to have a fixed set of allowed operations.
On current quantum computers, a small group of gates are carefully calibrated regularly to maintain their accuracies.
Overall, it is generally hard to take advantage of quantum pulses with NISQ machines.
% The reasons are the high noise of quantum hardware and the exceptional overhead of the pulse generator.

% WHY PULSES WORK FOR VQAs 
\noindent\textbf{Why Pulses for VQAs?} 
Situations differ slightly in the case of VQAs due to its variational characteristics.
During the ``training'' process, the parametric circuits of VQAs are updated iteratively.
It is now unimportant whether the controls are accurately implemented on quantum hardware as long as the parametric circuits can reach the desired states.
Instead of using gate-level compilation or QOC, we can implement quantum algorithms at the ``native-pulse level''.
At the native-pulse level, we can directly manipulate the native pulses that are supported by the quantum hardware.
This paradigm change grants more fine-grained control and thus making it possible for better performance, scalability, and robustness. 
A recent work~\cite{gokhale2020optimization} has shed light on the feasibility and potentials for such a paradigm change. 
However, the research on pulse level optimization is still in its infancy. 
The capability of quantum pulses has not been fully explored, nor are they already robust or scalable on NISQ machines. 
As summarized in Table~\ref{table_comp}, many critical issues remain unsolved.
% Existing works, for instance, do not consider noise or system models of NISQ devices.
% Most of these pulses are therefore incompatible with NISQ machines.

%OUR FRAMEWORK
To tackle these challenges, we propose \name, a native-pulse ansatz generator for VQAs.
\name~is the first to demonstrate the feasibility of native-pulse ansatz on NISQ machines.
% In \name, we use native pulses to construct the variational part of the circuits, which is often referred to as ansatz.
Instead of using rotation gates, we directly use the pulses that are natively supported by the quantum processors.
Compared with QOC, \name~has fewer parameters and can be easily deployed onto NISQ machines, while QOC with realistic system models requires huge computation resources.
On the other side, \name~is superior to gate-based methods, since \name~drops the abstraction layer of native gates and results in less circuit latency. 
%ADVANTAGES of OUR PAPER
We provide results from IBM's superconducting quantum computers~\cite{mckay2018qiskit}, while previous pulse-level works are only evaluated on simulators.

% CONTRIBUTIONS
\noindent{\bf Contributions.}
The goal of this paper is to construct native-pulse ansatz for VQAs and demonstrate in both simulators and NISQ machines.
The major contributions of \name~include: 

\begin{itemize}[leftmargin=*]
    \item {\bf Native-pulse ansatz.} 
    Our native-pulse ansatz is derived from native pulses that are extracted from quantum backends.
    In this way, we ensure that pulse ansatz is compatible with quantum hardware.

    \item {\bf Progressive learning. } 
    In \name, a non-gradient optimizer is employed.
    We provide a progressive way to ``grow'' our native-pulse ansatz in order to maintain a reasonable size for the parameters handled by the optimizer.
    New pulse blocks with zero amplitudes are appended at different ``steps'' of \name.
    This prevents the appended pulses from abruptly changing the ansatz circuits' overall unitaries.
    
    \item {\bf Results from NISQ machines. } 
    Experiments are conducted on simulators and NISQ machines. 
    The results show that the native-pulse ansatz outperforms the gate ansatz for VQA tasks in terms of accuracy and latency.
    
    % With \name, we can reduce decoherence error because of the short duration of \name.
    
    \item {\bf Exploration on frequency tuning.}
    We explore possible benefits of tuning pulse frequencies on transmons.
    Experimental results show that pulse frequency can be an extra degree of freedom for native-pulse ansatz.
    
\end{itemize}

\noindent\textbf{Evaluation Highlights.} 
Six NISQ machines are used to evaluate \name.
We achieve latency reductions of up to \textbf{97.3\%} compared to baselines. 
Accuracies up to \textbf{99.895\%} are attained for small-size VQE tasks.

\section{Background}
\label{sec2}

\subsection{Variational Quantum Algorithms (VQAs)} 
The variational quantum eigensolver (VQE) is one of the most popular and promising VQAs.\
VQE is employed primarily to solve the ground and low-excited states of quantum systems. It has important applications in quantum many-body physics, quantum chemistry, and other fields~\cite{kandala2017hardware, peruzzo2014variational}.
% It calculates Hamiltonian ground state energy using the variational approach.
% Given a Hamiltonian $H$ and a trial wavefunction $|\varphi\rangle$, the ground state energy $E_0$ is bounded by:
% \begin{equation}
    % E_0 \leq \frac{\langle\psi|H|\psi\rangle}{\langle\psi|\psi\rangle}
% \end{equation}
% The accuracy of the classical calculation method is restricted by the amount of computation required to model the exponentially growing Hamiltonian.
% VQE, on the other hand, can model complex wave functions in polynomial time.
VQE has also been shown to be noise resistant in NISQ devices~\cite{wang2019accelerated}.
The Quantum Approximate Optimization Algorithm (QAOA) is a VQA that solves combinatorial optimization problems with sub-optimal solutions~\cite{guerreschi2019qaoa}. 
% The ability to rapidly find the optimal variational parameters for generally shallow quantum circuits is essential for demonstrating the potential quantum advantages of NISQ devices.
Quantum neural network (QNN) is a model of quantum machine learning (QML) that use ansatz circuits to extract features from input data, followed by complex-valued linear transformations.
QNN has great potential for applications in QML~\cite{biamonte2017quantum}, quantum simulation, and optimization~\cite{moll2018quantum}.

\subsection{Quantum Optimal Control} 
Assuming we have a closed quantum system, the Hamiltonian of the system is provided by
\begin{equation}
    H(t) = H_0 + \sum_{j=1} u_j(t)H(j)
\end{equation}
where $H_0$ is the drift Hamiltonian, $H(j)$ is the control Hamiltonian, and $u_j$ is time-dependent control signal. 
The Schrödinger's equation governs the system's dynamics:
\begin{equation}
     \frac{d}{dt}|\psi\rangle = -iH(t)|\psi_0\rangle
\end{equation}
where $\psi_0$ denotes the system's state at time $t=0$.
Quantum optimum control (QOC) can be used to calculate $u_j(t)$ which corresponds with control signals.
With QOC, we can transform one quantum state into a desired state.
% The QOC generators create pulse signals with pulse simulators such as Qutip~\cite{johansson2012qutip} or Qiskit~\cite{mckay2018qiskit}. 
Typically, a cost function is specified as the fidelity of generated pulses, which measures the difference between the simulated unitary matrix and the desired unitary matrix.
Algorithms like GRAPE~\cite{de2011second} and CRAB~\cite{caneva2011chopped} have been developed to solve QOC problem.

\subsection{Quantum Pulse Learning} 
Another method of generating pulses is to parameterize the quantum pulses and then optimize the parameters.
We refer to such process as quantum pulse learning.
To comprehend the trainable components of control pulses for superconducting quantum computers, the Hamiltonian of transmon can be defined as follows:
\begin{equation}
\begin{aligned}
H = \sum_{i=0}^{1} (U_i(t)+D_i(t)) \sigma_i^{X} +
\sum_{i=0}^{1} 2\pi \nu_i (1-\sigma_i^{Z})/2 \\+
\omega_B a_B a^{\dagger}_B +
\sum_{i=0}^{1} g_i \sigma_i^{X} (a_B + a_B^{\dagger})
\label{eq4}
\end{aligned}
\end{equation}
$D_i(t)$ and $U_i(t)$ are two major terms that are determined by the pulse learning, they represent modulated signals as shown in Equation~\ref{eq5}. 
They are obtained by mixing a local oscillator with control signals.
$\sigma_X$, $\sigma_Y$, and $\sigma_Z$ are Pauli operators. 
$\nu_i$ is the frequency of qubit $i$, $g_i$ is the coupling strength between qubits, $\omega_B$ is the frequency of control buses, $a_B$ and $a^{\dagger}_B$ are the ladder operator for control buses. 

\begin{equation}
\begin{aligned}
D_i(t) = Re (d_i(t)e^{iw_{d_i}t})\\
U_i(t) = Re [u_i(t)e^{i(w_{d_i} - w_{d_j}) t})]
\label{eq5}
\end{aligned}
\end{equation}
where $d_i(t)$ and $u_i(t)$ are the signals of qubit $i$ on drive channel and control channel. 
Since pulse learning adjust $d_i(t)$ and $u_i(t)$, $D_i(t)$ and $U_i(t)$ are changed accordingly. 
Consequently, the drive Hamiltonian is updated~\cite{liang2022variational}.
In this way, we can manipulate the quantum system with control signals.
%Numerous publications have examined how to benefit from such scheme.
%For example, Ctrl-VQE~\cite{meitei2021gate} propose to optimize pulse shapes for the state preparation.
%Since these approaches do not target NISQ machines, they are subject to the vulnerabilities outlined in Section~\ref{sec1}.

% \subsection{Current NISQ machines}
% % HUAFEI TODO: 
% Different material systems have been explored to implement qubits and quantum gates.
% Material systems include neutral atoms~\cite{bluvstein2023logical}, trapped ions~\cite{paul1990electromagnetic}, optical lattices~\cite{grimm2000optical}, nuclear magnetic resonance (NMR)~\cite{vandersypen2001experimental}, diamond~\cite{neumann2008multipartite}, and superconducting circuits~\cite{clarke2008superconducting}.
% % Superconducting circuit is one of the leading candidates to build a quantum computer to perform the computations that are beyond the reach of classical computers. 
% In superconducting quantum computers, quantum gates are implemented by driving the target qubit through the microwave pulses\cite{alexander2020qiskit}. 
% % For example, the Rx and Ry rotation gates can be implemented by sending modulated microwave signals onto transmons. 
% For example, Derivative Removal by Adiabatic Gate (DRAG) pulses are used to implement single-qubit operations on transmons~\cite{alexander2020qiskit}.
% Due to the imperfect implementations and various sources of noises, efficient and precise pulse controls remain an open question, which also provides more optimization opportunities both on software and hardware levels.
\section{Related Work}
\label{sec7}
\textbf{Pulse Learning Approach: } 
Ctrl-VQE~\cite{meitei2021gate} tweaks the pulse shapes to perform state preparation. 
The methods are evaluated with Qutip~\cite{johansson2012qutip} pulse simulator.
And the results demonstrate that the total time required for state preparation is greatly reduced. 
VQP~\cite{liang2022variational} uses pulses as basic components to build the QNN ansatz and exhibits latency advantages over gate-based QNN on a two-class image classification task.
The experiments are conducted on Qiskit~\cite{mckay2018qiskit} pulse simulator.
VQOC~\cite{de2022pulse} presents a mathematical formalism of optimal control, which acts on pulse optimization for VQA tasks.
Their method is similar to Ctrl-VQE~\cite{meitei2021gate}, but they take advantages of neutral atom's properties.
The evaluations are also performed on simulator. 
These previous works are exploratory and rely on classical simulations of small quantum systems~\cite{meitei2021gate, liang2022variational}. \name, on the other hand, provides results from NISQ machines. Moreover, Ctrl-VQE is only designed with the consideration for single qubit pulse, whereas, \name~ both consider single qubit pulse and two qubit pulse.

\textbf{Ansatz Architecture Search: }
Hamiltonian simulation plays an important role in simulating quantum systems~\cite{ peruzzo2014variational,kandala2017hardware}. 
VQAs can be deployed to perform Hamiltonian simulation. 
And previous works~\cite{kandala2017hardware} pointed out that the choice of ansatz is important for VQAs.
The conventional approaches of choosing ansatz depends heavily on the applications. ~\cite{peruzzo2014variational,o2016scalable} are specially designed for VQE tasks. The unitary coupled-cluster singles and doubles (UCCSD) is still the golden standards for VQE ansatz.
QAS~\cite{du2022quantum} proposes a noise-aware scheme to search for ansatz structure.
The robustness to noise is demonstrated on simulators.
QuantumNAS~\cite{wang2022quantumnas} presents a comprehensive framework for noise-adaptive co-search of the ansatz. 
Hardware topology is considered during the search algorithm.
QuantumNAS validates their methods with evaluations on NISQ machines.
\begin{figure}[t]
\centering
\includegraphics[width=0.98\linewidth]{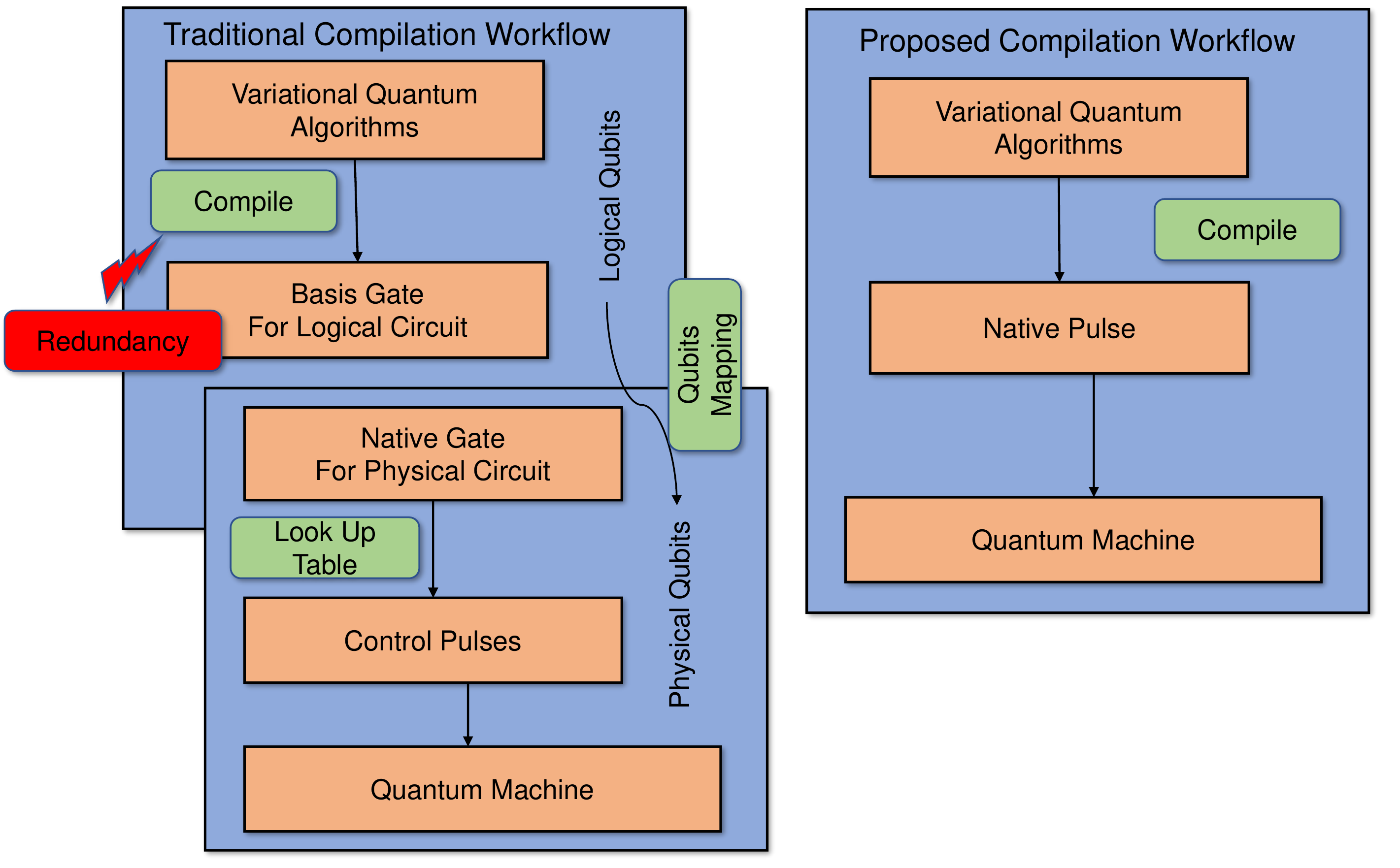}
\caption{Comparison between compilation process for gate level and pulse level. Gate-level workflow consists of several layers and introduces redundancy, pulse-level workflow consists fewer layers that can provide reduction of circuit latency.}
%\vspace{-5mm}
\label{compiler}
\end{figure}

\begin{figure}[t]
\centering
\includegraphics[width=0.98\linewidth]{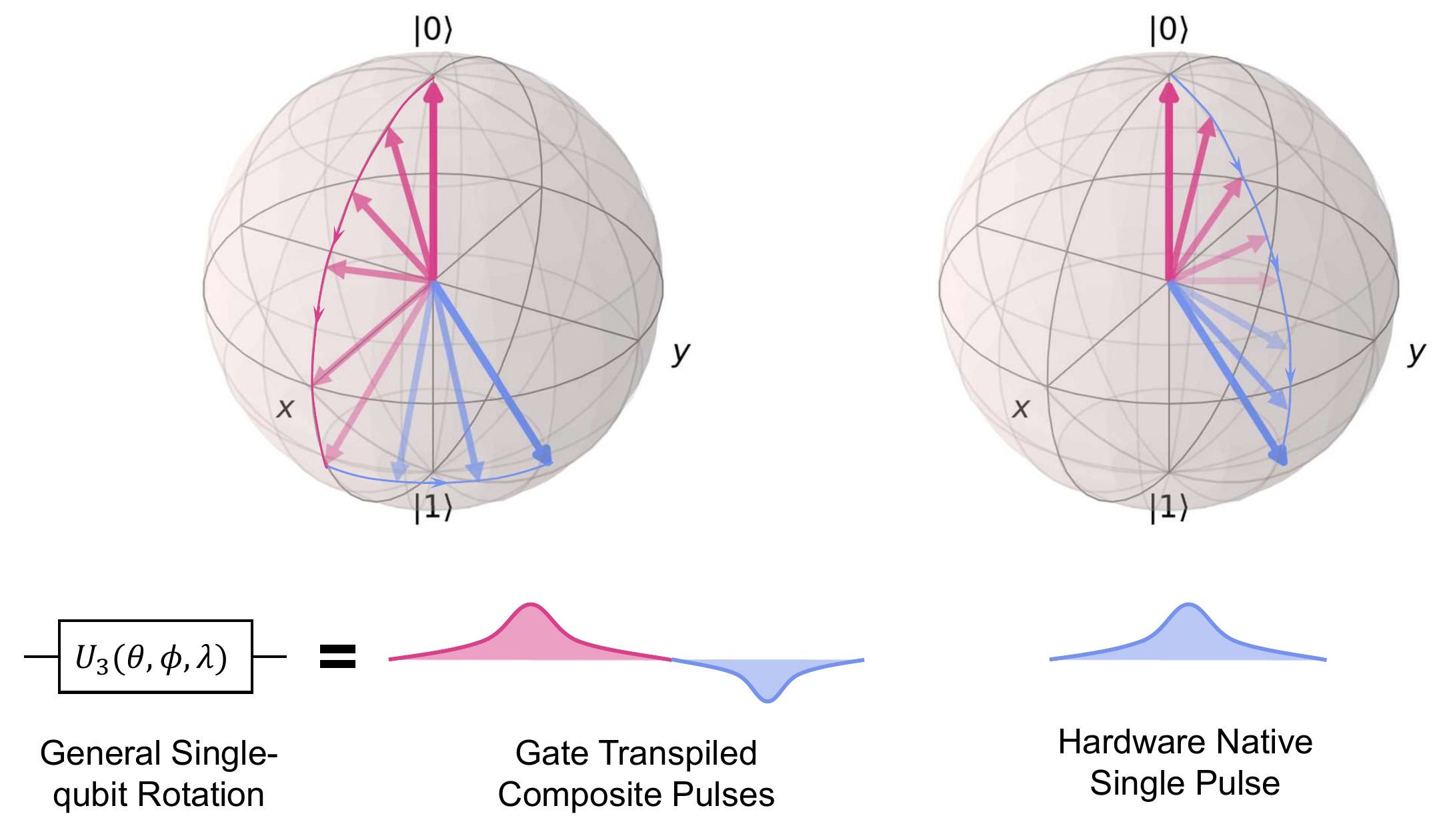}
\caption{
Illustrations of the redundancy introduced by current gate-level compilation workflow. 
In the case of gate-level compilation, pulses with fixed parameters are inserted to first rotate the qubit to another axis.
With pulse-level controls, we can avoid such redundancy and reduce latency without loss of capability to explore the Hilbert space.}

\vspace{-2mm}
\label{bloch}
\end{figure}

\textbf{Pulse Gate Compilation: }
\cite{gokhale2020optimized} proposes a new compilation paradigm, based on the OpenPulse interface for IBM quantum computers. 
This work achieves lower error rates and shorter execution times in comparison with traditional gate-based compilation methods.
Their technique is bootstrapped from existing gate calibrations, thus their pulses are in a simple form.~\cite{shi2019optimized} designs a general quantum compilation method to integrate multiple operations into larger units and then achieve high efficiency by optimizing this aggregation and creating customized control pulses.

\section{Motivation}
\label{sec3}

\subsection{Deficiencies of Gate-level Compiler}
Figure \ref{compiler} illustrates the compilation of a quantum program from high-level programming languages to the physical signals on quantum machines.
% Variational quantum algorithms use parametric gates in the standard compilation cycle.
Redundancy is introduced by the existing implementation of rotation gates on quantum hardware.
For example, we want the qubit to evolve to the point on the Bloch sphere as indicated in Figure \ref{bloch}.
In order to reach such state, a basis transformation, a phase shift, and a second basis transformation must be performed.
This example illustrates the redundancy when quantum circuits are compiled into native gates.
\revise{On the other hand, a single Derivative Removal by Adiabatic Gate (DRAG) pulse is capable of implementing an arbitrary single-qubit rotation.}
We propose to bypass the abstract layer of native gates and use native pulses directly as parametric elements in the ansatz.
\revise{In this way, this approach results in a pulse circuit with reduced latency and increased number of parameters, enhancing its flexibility.}

\subsection{Drawbacks of Optimal Control Pulses}
In addition, attempts have been made to optimize the quantum pulse generator.
For example, ~\cite{gokhale2019partial} proposes to use quantum optimal control to generate pulses for given unitaries.
The applications of quantum optimal control are severely constrained by the excessive cost of pulse generators and the complexity of NISQ devices' system models.
\red{
Besides, QOC is usually adopted when a target unitary or state transition is known. Hence, it is incompatible with VQAs, which lack target unitaries.
In addition, QOC's gradient-based approach with back propagation cannot be used, because gradients for pulses are not accessible on NISQ machines.
}
% As a result, the \name~must deal with non-gradient optimizers.
% Non-gradient optimizers, on the other hand, have a high computational cost and a substantial degree of uncertainty when dealing with high-dimensional spatial parameters.
% This is because a large number of samples must be frequently pre-set or executed throughout the processing procedure.
% Because pulses contain more parameters than gates, it is inefficient to train the whole pulse ansatz with a non-gradient optimizer.

\subsection{Potentials of Pulse Ansatz} 
As the illustration in Figure \ref{bloch}, if we can directly control pulses, we can eliminate some redundancy. 
% This allows us to reduce the program's overall latency.
% Specifically, decoherence errors decrease.
\red{
Using pulses rather than gates can result in a lack of calibrations, which may result in inacurate operations. 
However, for VQAs, the parameterized pulse ansatz can be automatically adjusted for errors including under-rotation and over-rotation.
}
Besides amplitudes of native pulses, the frequencies of qubit channels can also be tuned.
In conclusion, pulse ansatz offers more degrees of freedom, enabling us to search for desired states with substantially shorter pulse latencies.
\begin{figure*}[t]
\centering
\includegraphics[width=0.97\linewidth]{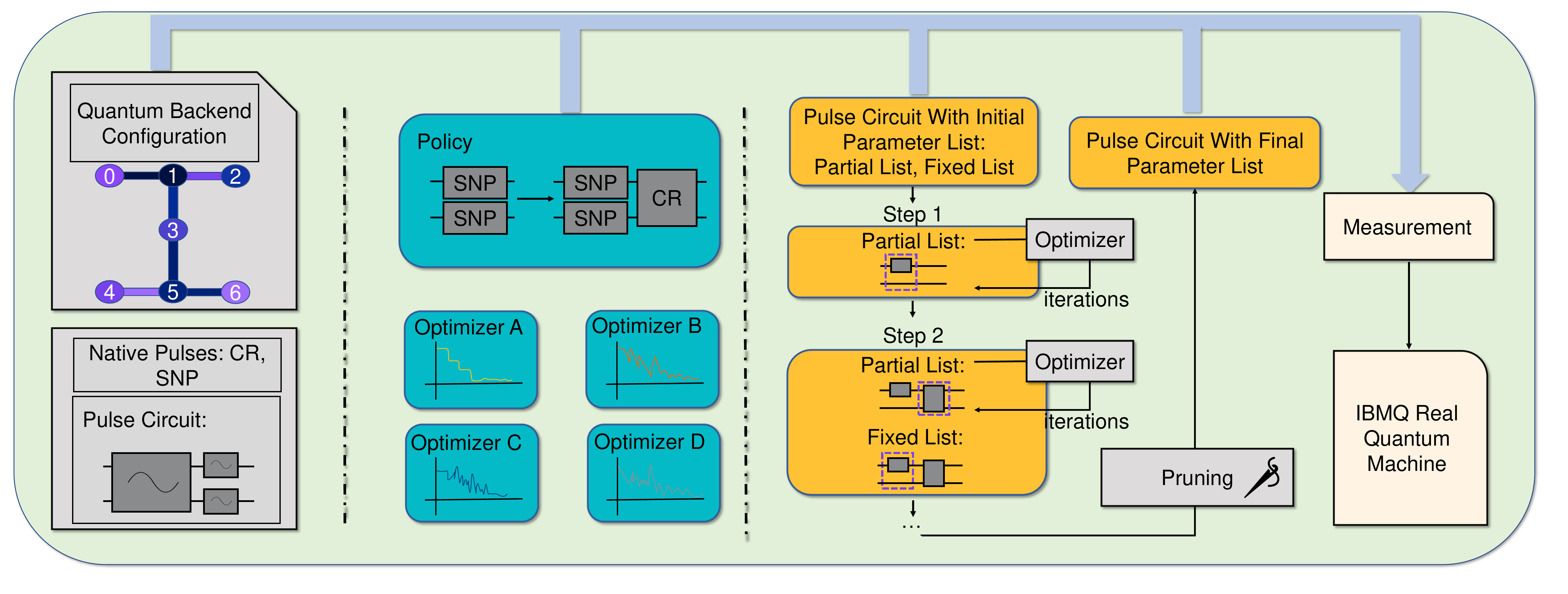}
\caption{\revise{The overview of design and implementation of \name. The proposed pulse ansatz is composed of single-qubit native pulses (SNP) and cross-resonance pulses (CR). 
During the training process, the ansatz is ``grown'' after each step. In this progressive way, older parameters from the last step are unchanged as a ``fixed list'', while newer parameters added in this step are updated by the optimizer through optimization iterations as a ``partial list''.
}
}
%\vspace{-3mm}
\label{overview}
\end{figure*}
\section{Overview of \name~Framework}
\label{sec4}

In the proposed framework for native-pulse ansatz, the pulse parameters determine the drive Hamiltonian in Equation~\ref{eq4}.
% To ``train'' the pulse ansatz on NISQ machines, however, gradients are not available. %using back propagation (BP).
However, gradients are not available as we train the pulse ansatz on NISQ machines.
%For a given unitary matrix, gradients can be computed using the QOC pulse generation method, but not the pulse learning method.
Therefore, we have to use a non-gradient optimizer to train the parameters in the ansatz. 
% In high-dimensional space, however, none of the optimizers can gain an advantage over a random parameter search.
To mitigate the drawbacks of non-gradient optimizer, we offer a progressive method to generate our native-pulse ansatz.
Such progressive learning structure ensures that the optimizer's parameter dimensions do not exceed its capacity.
This machine-in-loop training method also makes the framework noise-resilient.
% Such a system is well-suited for pulse-level learning algorithms to address the noise issue in NISQ machines while minimizing the number of parameters handled by the optimizer.
We illustrate the workflow of the proposed framework \name~in Figure \ref{overview}.
Firstly, the configurations of the NISQ computers are extracted, which include information regarding qubit frequencies and the mapping between native gates and native pulses.
% Once we obtain the native pulses supported by the backends, we can generate native-pulse ansatz accordingly.
% The native-pulse ansatz is constructed from the extracted pulses' structures.
% The extracted pulses are compatible with quantum hardware, and we can adjust the amplitudes to achieve different functions.
After obtaining the native pulses from quantum backends, we can start to progressively construct our native pulse ansatz.
Inspired by the hardware-efficient ansatz (HEA), single-qubit gates are placed for each qubit and two-qubit gates are applied to all available connections.
The single-qubit pulses are derived from pulses that correspond with the Hadamard gate or Rx gate.
And two-qubit pulses are derived from CX gate or CR \rebuttal{pulse}.
CR \rebuttal{pulses} are preferred because they are the simplest pulses that enable the entanglement of two qubits.
Now, we have two types of pulse layers. 
One consists of single-qubit native pulses on all qubits, whereas the other consists of two-qubit native pulses on available connections.
The two types of layers are alternately inserted during the training process to help explore the Hilbert space.
To train the native-pulse ansatz, we use non-gradient optimizers.
% Although gradients cannot be derived from parameter-shift rules like the gate ansatz, non-gradient optimizers minimize the cost function using finite-difference approaches.
The incrementally constructed ansatz prevents non-gradient optimizers from failing to work with huge dimensions.

Simulators and NISQ machines are employed to evaluate our methods.
On the simulators, we present energy-distance curves for several molecules. The energy curves closely resemble those generated with the full configuration interaction (FCI) approach.
The results show that our pulse ansatz can approximate the lowest energy states of molecules with much shorter pulse latencies.
Our ansatz are also tested on NISQ machines, which have realistic noise and complex system model.
% With \name, we achieve greater performance than previous works.
In addition, we demonstrate that tuning frequencies on NISQ computers is beneficial.

\section{Design and Implementation Details} 
\label{sec5}
% \begin{table*}[t]
% \centering
% \renewcommand*{\arraystretch}{1}
% \setlength{\tabcolsep}{6.5pt}
% \footnotesize
% \caption{Comparison of trainability for different pulse circuits and gate circuits on ibmq\_jakarta.}
% \begin{tabular}{c|c|c|c|c|c}
% \midrule
% \textbf{Operations}       & \textbf{Circuit Level} & \textbf{Molecule Bond Length} & \textbf{Reference Energy} & \textbf{VQE ($H_2$) Result} & \textbf{Duration(on ibmq\_jakarta)}   \\ \midrule
% SNP & Pulse Circuit  &0.1\AA & 2.710H& 4.380H & 71.1ns\\ \midrule
% CR & Pulse Circuit  &0.1\AA &2.710H &2.927H & 163.6ns\\ \midrule
% SNP & Pulse Circuit  &0.75\AA &-1.137H&-0.549H & 71.1ns\\ \midrule
% CR & Pulse Circuit  &0.75\AA &-1.137H &-1.032H & 163.6ns\\ \midrule
% CR + SNP & Pulse Circuit &0.75\AA &-1.137H &-1.036H & 234.7ns\\ \midrule
% Two Gate Ansatz   & Gate Circuit  &0.75\AA   &-1.137H      & -0.534H & 341.3ns        \\ \midrule
% \end{tabular}
% \label{compare}
% \end{table*}
\subsection{Gate Ansatz versus Native-Pulse Ansatz}
In the proposed \name, we replace the basic element of the variational quantum circuit with a native pulse and use quantum pulses to build a native-pulse ansatz.
During the training process, parameters of native pulses are updated, thus obviating the requirement for decomposition into native gates.
To create a native-pulse ansatz, we intentionally make sure that the employed pulses are parametric pulses supported by the NISQ device.
% In the process, it is optional whether they correspond to a particular gate.
Since we use native pulses as building blocks for our ansatz, we can explore more available parameters that are allowed to be tuned.
In the case of gate-based ansatz, the parameters are limited to angles of rotation gates.
These gates are decomposed into native gates before being implemented with microwave pulses and phase shifts.
The microwave pulses are applied to qubits, while the phase shift is influencing on classical electronics.
When the angles for the rotation gates change, it changes the phase-shift operations. However, the ``redundant'' microwave pulses remain fixed.
Consequently, gate-based methods cannot take full advantage of parametric pulses.
So, we choose to directly adjust the parameters of the microwave pulses that are acting on qubits.
In this way, we introduce opportunities to contain more trainable pulses within the same circuit latency.
That is, the latency cost for each parametric pulse is reduced, making it possible to generate pulse ansatz with more parameters and less latency at the same time.
% As shown in Figure \ref{compiler}, we are able to reduce the number of abstract layers.
Another advantage of adjusting pulse parameters instead of gate angles is the mitigation of gate-to-pulse compilation noise. Moreover, for pulse-end cloud users, from scratch to build a pulse ansatz could greatly save the calibration cost which is pretty expensive.
To experimentally realize a continuous parametric gate such as RX($\pi$/4), the amplitude of $X_{\pi/4}$ pulse is set to be half of that of a $X_{\pi/2}$ pulse, i.e., $A_{\pi/4} = A_{\pi/2}/2$. 
However, this step introduces noise due to the nonlinearity in NISQ devices\cite{Gil_Lopez_2021}. 
\name~can directly tune the underlying pulse parameters and avoid such gate-to-pulse compilation noises. 

%In summary, the parameters of native-pulse ansatz include amplitudes and frequencies.
%Before deploying native-pulse ansatz, we did some analysis on the trainable parameters.
%First, in VQC, training of the parametric circuit is actually theoretically required to change the area of the pulse physically so that it can rotate to the target on the Bloch sphere. 
%but as we have already discussed, the current IBM superconducting quantum computer is mainly by adjusting the classical electronics on the phase to train the parametric circuit, which results in a large redundancy. 
%Thus, we choose to directly train the amplitude on the quantum pulses, thus directly causing the area of the physical pulse to vary, which can obtain efficiency. 
%We can save a lot of layers when we build the circuit based on native pulses as shown in Figure \ref{compiler}. 
%While not requiring the compilation of gates into pulses, it also gives a simpler representation of the overall circuit. 
%But even if these redundancies exist, if we have a perfect enough quantum computer that can support a quantum circuit of sufficient depth, then the trainability of the amplitude training of the pulse and the gate level training will be the same, which is like a person traveling to a destination, walking diagonally and walking straight before walking horizontally will eventually reach the destination, only the cost will be different.
\subsection{Analysis the Power of Native-Pulse Ansatz}
In the realm of quantum computation, the precision of operations, especially quantum gates, is quintessential.
\begin{figure}[t]
\centering
\includegraphics[width=\linewidth]{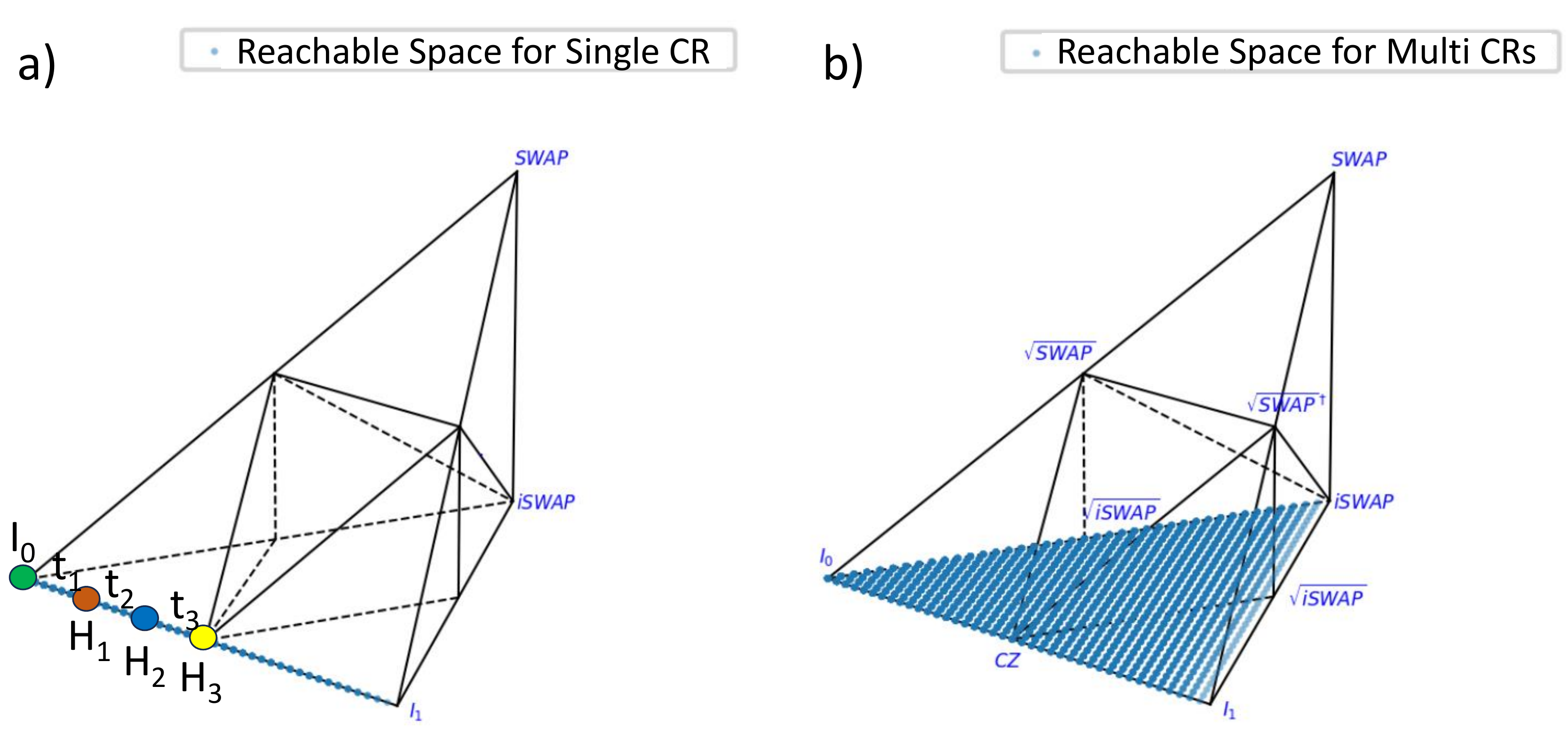}
\caption{\revise{Reachable space comparison on Weyl Chamber for a) Single-CR. b) Multi-CRs. While a single-CR pulse can only cover one dimension on the Weyl Chamber, multi-CRs cover one of the surfaces of the Weyl Chamber.}}
% \vspace{-1mm}
\label{weyl}
\end{figure}

\begin{figure}[t]
\centering
\includegraphics[width= 0.97\linewidth]{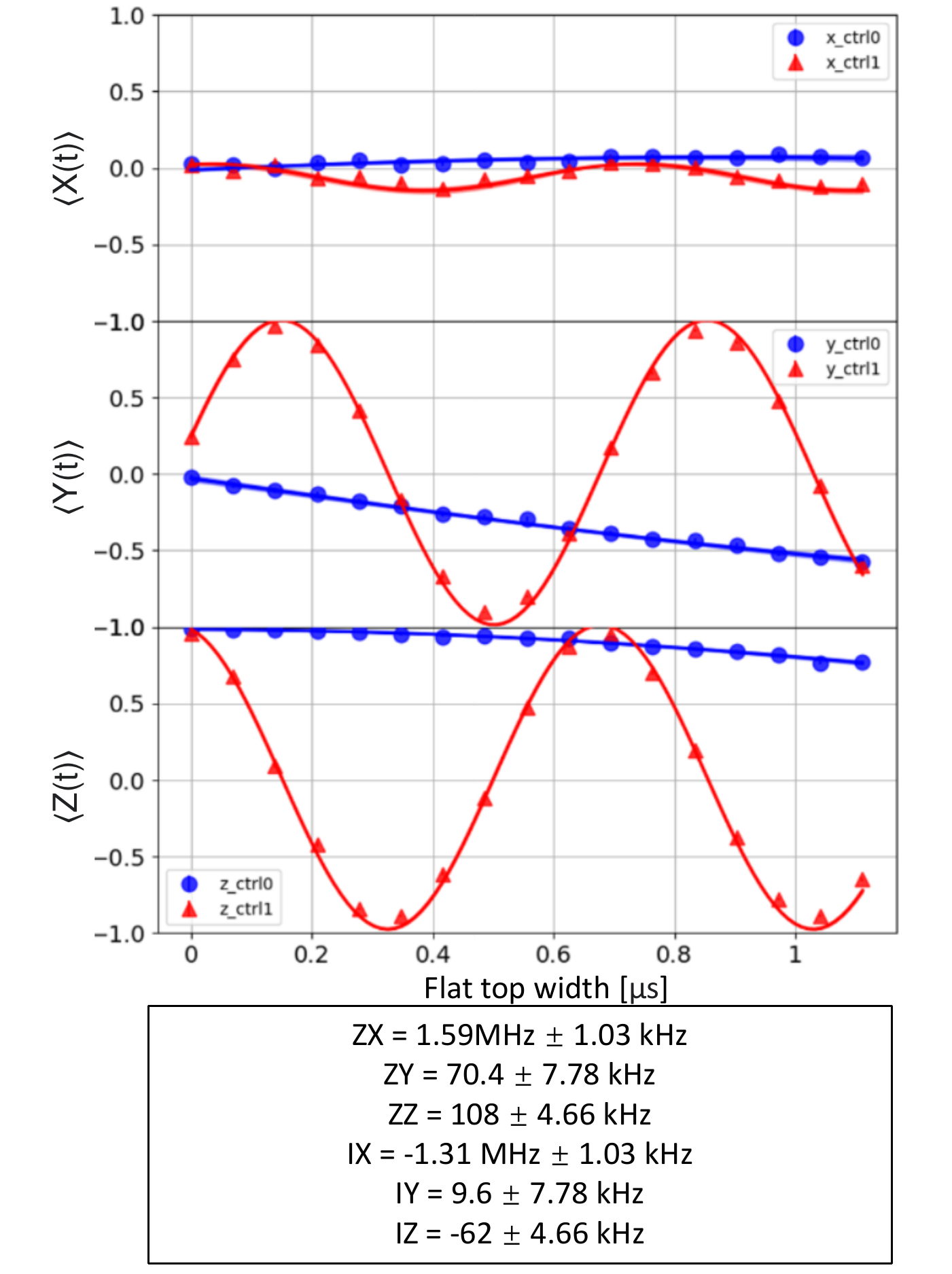}
\caption{Cross-resonance pulses have several components other than the intended ZX interaction. While this poses challenges for gate-level calibrations, it doesn't affect parametric pulses.}
% \vspace{-1mm}
\label{tomo}
\end{figure}
\noindent \textbf{Traditional Quantum Rotation Gates:}
Traditional quantum rotation gates such as \(RX\) and \(RY\) possess durations of \(320 \text{dt}\), where \(\text{dt}\) represents a standardized time unit equal to \(0.22\)ns. Predominantly, the training process for these gates is inclined towards modulating the angle of the quantum rotation gates. This modality implies that the underlying physical pulse remains unaffected, with adjustments predominantly focusing on the classical phase.

\noindent \textbf{Single Qubit Native Pulse:}
Transitioning our attention to the single qubit native pulse (SNP), its duration is notably \(160 \text{dt}\). Instead of merely modifying the classical phase, one can directly train on the parameters of the physical pulse. This alteration to the pulse parameters allows for an effective influence on the drive Hamiltonian, consequently crafting various operations. We propose that harnessing the power of the pulse ansatz in VQAs can transcend the traditional limitations of standard quantum gate operations.

\noindent \textbf{Cross-Resonance: }
However, one needs to approach this perspective with caution. When deconstructing a quantum circuit, it becomes evident that the influence of a single qubit gate or pulse in terms of the overall circuit duration is relatively minor. The true juggernauts are the two-qubit gates or pulses, which indisputably play the role of a major term in shaping the circuit's dynamics. In the sphere of superconducting quantum machines, the cross-resonance pulses (CR) stand as the minimal two-qubit operation. Given its prominence and efficiency, our choice naturally gravitated towards employing cross resonance as our primary two-qubit operation in the formulation of the native-pulse ansatz.

Through rigorous testing via uniformly random sampling on the Weyl chamber, we derived the gate coverage for both single CR and multi CR as shown in Fig. \ref{weyl}. Remarkably, the gate coverage of the single CR parallels that of the CPhase gate. Moreover, when the angle is set to \(\pi/2\), it is locally equivalent to either the CX or CZ gates. This local equivalence is particularly insightful, as gates that are locally equivalent can be interconverted through the addition of specific single-qubit operations. The analysis of multi-CR unveils a more comprehensive coverage, particularly spanning the entire base of the Weyl chamber. This extensive coverage encompasses a range of standard gates such as the B gate, iSwap, and several others. Interpreting this, when adopting CR as the two-qubit native pulse, we harbor the potential to not only emulate a multitude of standard gates but also venture into the realm of some nonstandard gates. To further elucidate the intricacies of the CR pulse, \revise{we provide the equation to describe the effective Hamiltonian of the CR pulse as below:
\begin{align}
            H &= Z \otimes A_2 + I \otimes B_2 \nonumber \\
            &= a_x \hat{Z} \hat{X} + a_y \hat{Z} \hat{Y} + a_z \hat{Z} \hat{Z} + b_x \hat{I} \hat{X} + b_y \hat{I} \hat{Y} + b_z \hat{I} \hat{Z}.
\label{crh}
\end{align}}
\revise{For example, we start with an initial state $t_0$ which is located at the point $I_0$ at the weyl chamber in Fig. \ref{weyl} (a).} \revise{Now, based on the strength of the Hamiltonian terms we measured from the Hamiltonian tomography experiment with a given pulse duration $t_1$, we can calculate the Hamiltonian based on Equation. \ref{crh}, once we obtain the Hamiltonian, we can further refer to the equation:}
\begin{equation}
U = e^{-iHt}
\end{equation}
to obtain the unitary gate or state evaluation results in matrix representation (here we take $t_0$ equals 70.4 ns and approximate the result to three decimals):
\[
\scriptsize
\begin{bmatrix}
    0.998 - 0.007j & -0.013 - 0.045j & 0 & 0 \\
    0.013 - 0.045j &  0.998 + 0.007j & 0 & 0 \\
    0 & 0 & 0.894 + 0.026j & 0.009 + 0.447j \\
    0 & 0 & -0.009 + 0.447j & 0.894 - 0.026j
\end{bmatrix}
\]

\begin{table*}[t]
    \centering
    \renewcommand*{\arraystretch}{1.4} % Adjusted for consistent row height
    \setlength{\tabcolsep}{9pt} % Adjusted for consistent column separation
    \footnotesize
    \begin{tabular}{|c|c|c|c|c|c|} % Added vertical lines for consistency
        \hline % Added top horizontal line
        \textbf{Operations} & \textbf{Circuit Level} & \textbf{Molecule Bond Length} & \textbf{Reference Energy} & \textbf{VQE (\(H_2\)) Result} & \textbf{Duration (on ibmq\_jakarta)} \\
        \hline % Added horizontal line after headings
        SNP & Pulse Circuit & 0.1\AA & 2.710H & 4.380H & 71.1ns \\
        \hline % Added horizontal lines between rows for consistency
        CR & Pulse Circuit & 0.1\AA & 2.710H & 2.927H & 163.6ns \\
        \hline
        SNP & Pulse Circuit & 0.75\AA & -1.137H & -0.549H & 71.1ns \\
        \hline
        CR & Pulse Circuit & 0.75\AA & -1.137H & -1.032H & 163.6ns \\
        \hline
        CR + SNP & Pulse Circuit & 0.75\AA & -1.137H & -1.036H & 234.7ns \\
        \hline
        Two Gate Ansatz & Gate Circuit & 0.75\AA & -1.137H & -0.534H & 341.3ns \\
        \hline % Added bottom horizontal line
    \end{tabular}
        \caption{Comparison of trainability for different pulse circuits and gate circuits on ibmq\_jakarta.}
    \label{compare}
\end{table*}
\revise{this indicates after implement the specific CR pulse with duration $t_1$, the Hamiltonian evolute to the point $H_1$ at Fig. \ref{weyl}. Repeat the process with time evolution $t_2$ and $t_3$, we can observe the evolution of the Hamiltonian on the Weyl chamber in Fig. \ref{weyl}.} This inherent flexibility augments the potential of the pulse ansatz in VQAs. By allowing exploration across multiple directions in the Hilbert space, we inherently boost the search capability within the design space.

\subsection{Parameters of Native-Pulse Ansatz}
\noindent\textbf{Tuning amplitudes.}
%In the NISQ machine, we have to face the challenge of limited decoherence time.
With native-pulse ansatz, we can tune parameters that are not accessible in gate-based ansatz.
For example, the amplitudes of the pulses on the drive and control channels are invisible to gate-level users.
% Here, we define two types of native pulses in our framework.
% Single-qubit native pulses (SNPs) are based on native pulses for one-qubit gates.
% And CRs) which is the most basic element that triggers the entanglement of two qubits.  
Both SNP and CR are initialized with zeros as their amplitudes. 
The reason is explained in section \ref{sub_ppl}.

% Moreover, in native-pulse ansatz there are parameters that cannot be tuned in VQC, and the first one to be analyzed and introduced into the \name~framework is the pulses on the control channel. 
% At gate level we current cannot realistically make any modifications to the multi-qubits gate, which is a problem. 
% Thus, we build pulses with the parameters refer to CR gate on pulse level and initial amplitudes to zero, and we call these pulses as two-qubit native pulse (TNP). 
% We further build pulses with the parameters refer to RX gate on pulse level also initial amplitudes to zero, these pulses determined as single-qubit native pulse (SNP).
% We believe it can be gainful by training the parameters of TNP on control channel, since TNP is a ``real'' basis unit on device level. 

% \textbf{Theoretical Analysis and Experimental Results of Tuning Multi-Qubits Gate.} 

We can see from Eq. \ref{eq4} and \ref{eq5} that when we change the parameters of the pulses, it changes the strength of the signal on the control channel of $d_i(t)$, which eventually changes the drive Hamiltonian. 
Experimental results presented in Table \ref{compare} confirm the advantages of pulse ansatz over gate ansatz. 
For experiments in Table \ref{compare}, the target is to solve the ground state energy of $H_2$ with VQE on the NISQ machine ibmq\_jakarta.
By comparing the results of SNP, CR, CR+SNP, and the two-gate ansatz, we demonstrate that the pulse ansatz provides a better energy value for the task in 31.2\% less duration.
The purpose of comparing the two-gate ansatz and the CR+SNP pulse circuit is to highlight the advantages of native-pulse ansatz construction, as CX and CR are the simplest 2-qubit operations on the gate-level and pulse-level, respectively.

% By comparing the experiment results of two SNPs separately addressed on two qubits and ``TNP + SNP'' pulse circuits, we can observe that the TNP have trainability, because the results of TNP + SNP have a clear advantage. We then introduce the experiments for a two gate ansatz.
% By comparing the results of TNP + SNP pulse circuit and two gate ansatz, we can see that the pulse circuit achieves an advantage of nearly 31.2\% in duration with achieve similar performance. 
% The reason for comparing the two-gate ansatz and TNP+SNP pulse circuit is to further emphasize the importance of native-pulse ansatz construction. 
% Our two-gate ansatz include CX gate, the smallest basis gate that can be compiled by the traditional compiler to achieve entanglement, while TNP is the smallest native pulse that we can compile in the pulse circuit to achieve entanglement. 

\begin{figure}[t]
\scalebox{1.15}{
\Qcircuit @C=1.0em @R=0.2em @!R { \\
	 	\nghost{{q}_{0} :  } & \lstick{{q}_{0} :  } & \ctrl{1} & \gate{\mathrm{H}} & \gate{\mathrm{H}} & \ctrl{1} & \meter & \qw & \qw & \qw\\
	 	\nghost{{q}_{1} :  } & \lstick{{q}_{1} :  } & \targ & \qw & \qw & \targ & \qw & \meter & \qw & \qw\\
	 	\nghost{\mathrm{{c} :  }} & \lstick{\mathrm{{c} :  }} & \lstick{/_{_{2}}} \cw & \cw & \cw & \cw & \dstick{_{_{\hspace{0.0em}0}}} \cw \ar @{<=} [-2,0] & \dstick{_{_{\hspace{0.0em}1}}} \cw \ar @{<=} [-1,0] & \cw & \cw\\
\\ }}
\includegraphics[width=0.98\linewidth]{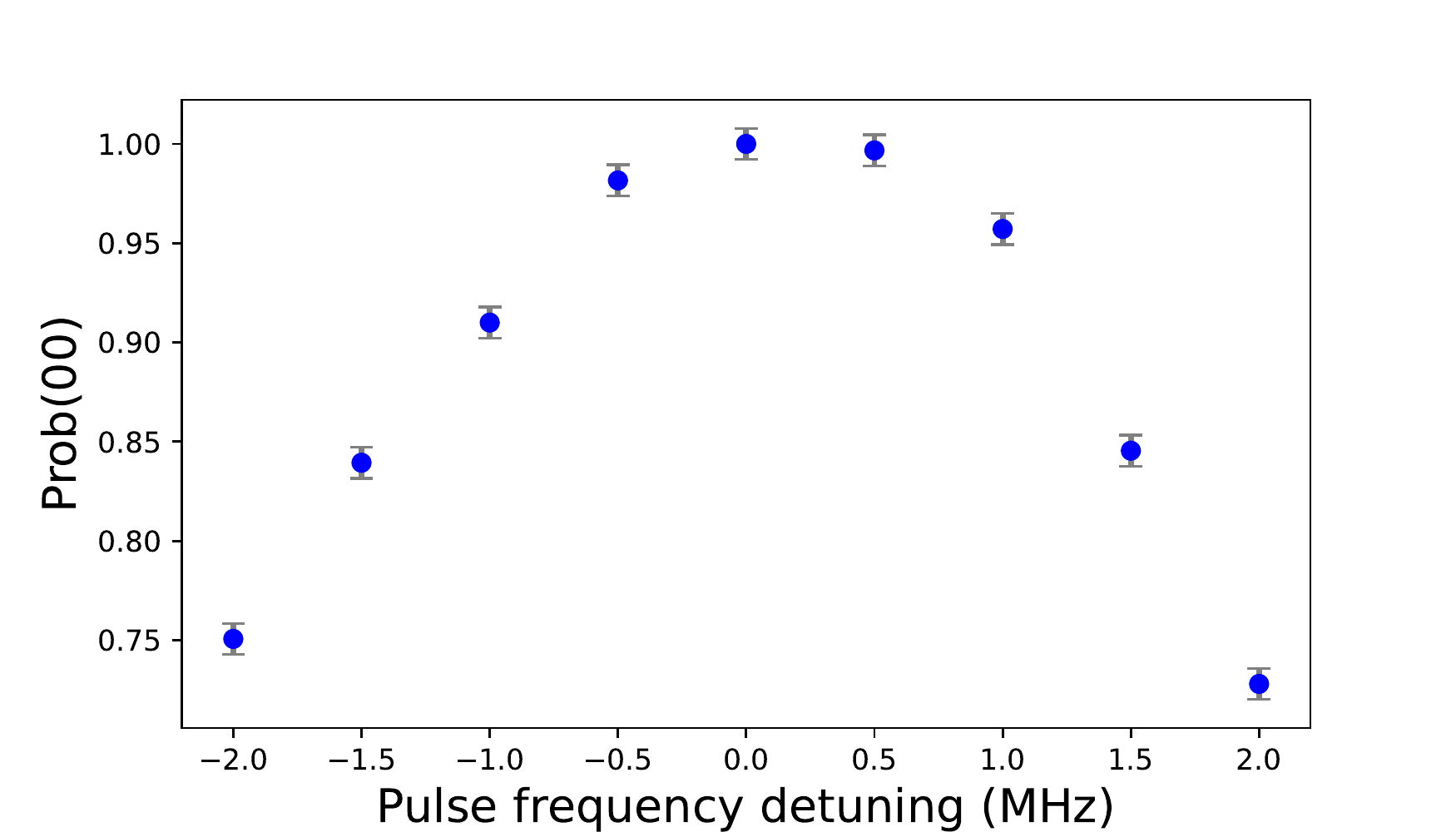}
\caption{The circuit is implemented by frequency-tunable pulses. 
\red{The high probability of getting $|00\rangle$ state for the zero detuning indicates that parametric pulse operations are physically feasible and can be implemented with high fidelity. 
}The changes of final state under different detuning indicate that the variation of pulse parameter indeed physically changes the corresponding quantum operations. Note that here we apply the measurement error mitigation technique \cite{https://doi.org/10.48550/arxiv.2007.03663} which reduces the impact of readout errors.}
%\vspace{-2mm}
\label{verify_physical_pulse}
\end{figure}

\begin{figure}[t]
\centering
\includegraphics[width=0.98\linewidth]{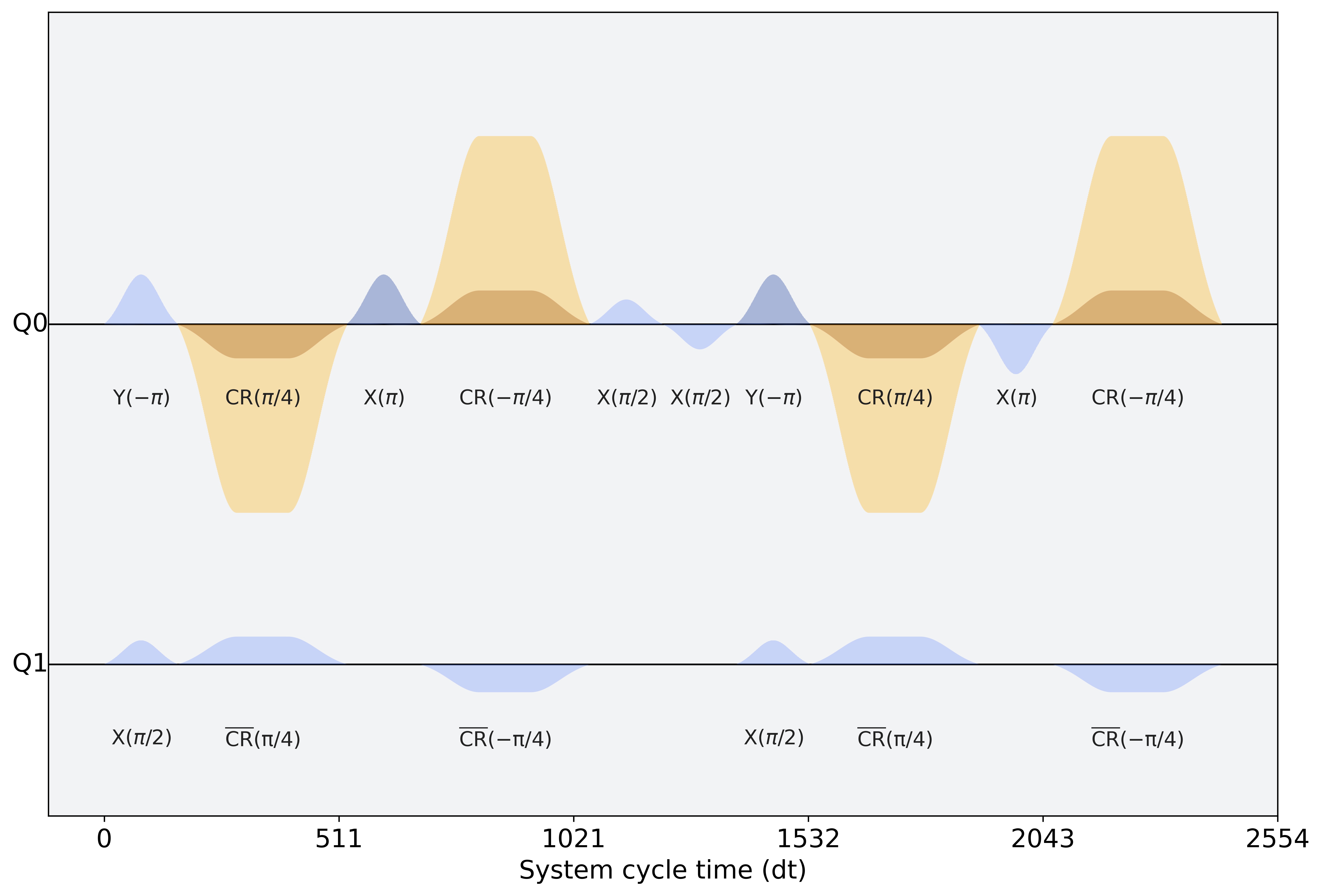}
\caption{\revise{Pulse schedules of ``CX+H+H+CX'' circuit. D0 and D1 are the drive channels on the first and the second qubits. They control the transmition from input signals to gate operations. U0 is the control channel which provides supplementary control over the qubit to the drive channel. The two qubit basis gate \rebuttal{CX} is realized through two $\pi/4$ cross-resonance pulses plus one $\pi/2$ X pulse. Sinlge qubit gates are first XY decomposed and then implemented by XY pulses. \cite{PhysRevA.96.022330}}}
% \vspace{-1mm}
\label{CXHHCXpulse}
\end{figure}

\noindent\textbf{Tuning frequencies.}
Clifford and T gates are universal to perform arbitrary quantum operations\cite{PhysRevA.57.127}, so gate calibrations mainly fine-tune a discrete set of $\pi/2$ and $\pi$ pulses corresponding to Clifford operations\cite{Theis_2018}. In the progressive pulse learning protocol, pulse parameters are continuously varied, so it is necessary to verify that parametric pulses still produce physical quantum operations, i.e., pulses correspond to quantum operations and can be implemented with a high fidelity\cite{RevModPhys.84.621}. 
We demonstrate the performance of pulse implementation by running a pulse version of the gate sequence $(CX+H+H^\dag+CX^\dag)$ under different frequency detuning. If the pulse block $(CX +H)$ is physically feasible to realize, when pulse detuning is zero, the circuit should produce a final $|00\rangle$ state with high probability. Experiment results in Figure \ref{verify_physical_pulse} show a near-to-one \rebuttal{probability} (00) readout result for the zero detuning case and thus prove the feasibility of parametric pulse operations. The high-fidelity results also indicate that \name~does not introduce additional noise and can be authentically implemented. 

In addition, the \rebuttal{results with frequency detuning} verify that the magnitude of the detuning range we choose ($\approx 2 MHz$) is wide enough such that the variation of pulse parameters can lead to the obvious changes in the corresponding operation.\rebuttal{The frequency could lead to leakage outside of the computational sub-space that may allow the optimizer to take short-cuts for better convergence~\cite{egger2023study}.} \rebuttal{In conclusion, frequency detuning with proper range is beneficial for VQAs.}
\begin{figure}[t]
\centering
\includegraphics[width=0.99\linewidth]{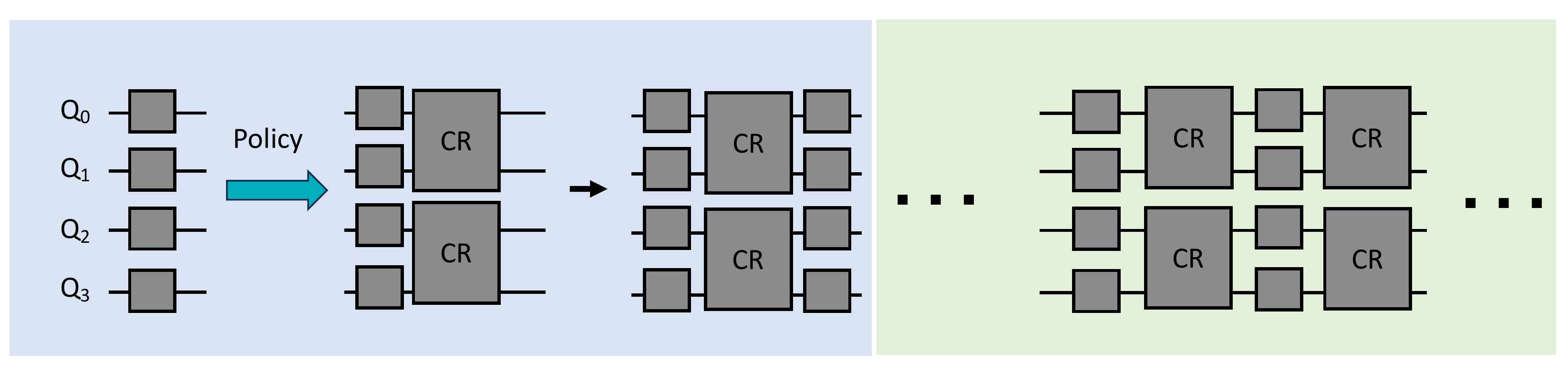}
\caption{\red{Examples of a policy to grow the native-pulse ansatz. The policy grows the CR and SNP in different steps. CR is designed to induce local entanglement.}}
%\vspace{-5mm}
\label{policy}
\end{figure}

\begin{figure}[t]
\centering
\includegraphics[width=0.95\linewidth]{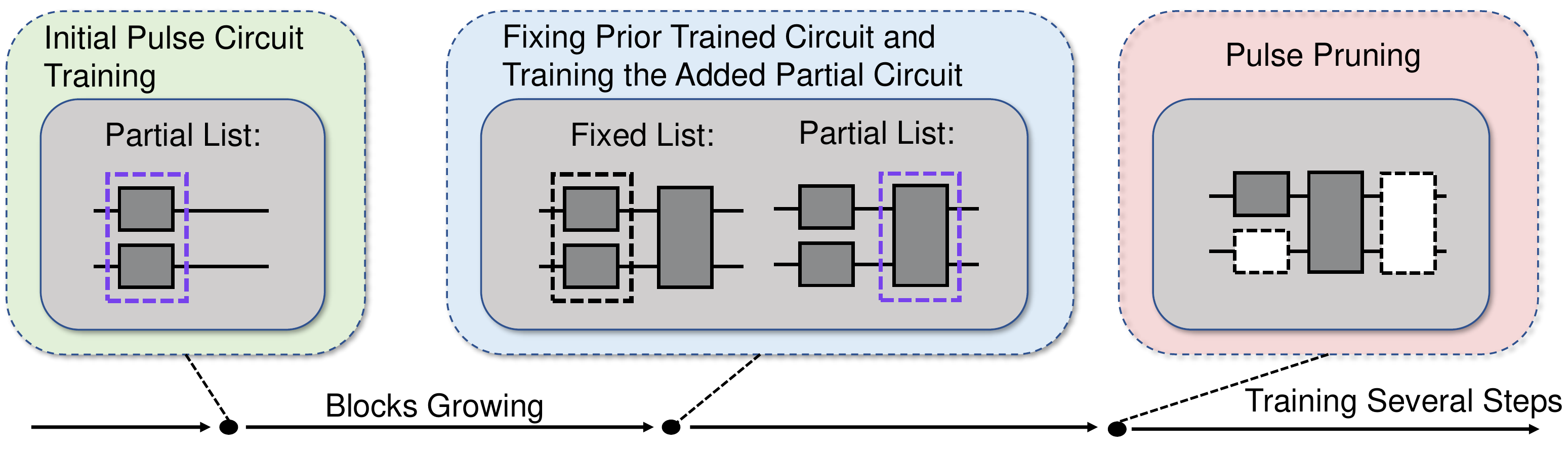}
\caption{During the training phase, we can determine which part of the parametric pulse is trained.
The complete list of parameters comprises a ``fixed list'' and a ``partial'' list. 
Pruning approaches were provided during training to further simplify the pulse ansatz.
}
%\vspace{-3mm}
\label{progressive}
\end{figure}

\subsection{Ansatz Construction}

VQAs usually consist of parametric circuits with a fixed structure. 
The major focus in the current VQA research community is to find a way to guide the construction of an ansatz. 
Traditional ways are mostly physics or chemistry-inspired, which can be cost-inefficient. 
For instance, to achieve certain accuracy for algorithms with time-evolving blocks such as QAOA, a few trotter steps are sufficient\cite{https://doi.org/10.48550/arxiv.2205.02520}. 
However, the structured ansatz is not adaptive, and it may be wasteful when we aim to achieve a less demanding accuracy. 
In comparison, our progressive method generates pulse ansatz adaptively.
The corresponding advantage is that the circuit depth is tailored for arbitrary desired accuracy and does not cause any experimental resource overhead. 

%begin{figure*}[t]
%\centering
%\includegraphics[width=1.0\linewidth]{figures/trendsforall.pdf}
%\caption{\red{Energy convergence trends for a) Ctrl-VQE~\cite{meitei2021gate}. b) \name. c) ``Brute-force'' pulse ansatz training. These three are identical in duration and are optimized with the same non-gradient optimizer. The final accuracies are comparable, whereas the NAPA converges with the fewest iterations.

%\vspace{-3mm}
%\label{trendsall}
%\end{figure*}

\subsection{Progressive Pulse Learning}
\label{sub_ppl}

\red{
Figure \ref{policy} gives an example of a policy to grow the native-pulse ansatz. 
This policy is inspired by the HEA proposed by~\cite{kandala2017hardware}.
Similar to HEA, single-qubit native pulses are applied to all qubits, while two-qubit native pulses are applied to available connections between qubits.
\revise{In this way, we create a pulse ansatz that parallels the gate-level HEA structure. This approach allows for a similar arrangement of multiple qubits as in HEA.}
The difference between \name~and HEA is whether a progressive policy is adopted.
The progressive manner of \name~can limit the number of parameters held by the optimizer.
Figure \ref{progressive} gives more details on our progressive policy. 
\revise{Importantly, the combination of SNPs and CRs is capable of covering the universal quantum gate set. SNPs and CRs are more essential components of the current basis gates used in quantum computing. For single-qubit operations, any arbitrary rotation can be accomplished using a single native pulse. Similarly, two-qubit operations are implemented through cross-resonance gates. For instance, CNOT gates can be realized using echoed cross-resonance gates. Thus, the combination of SNPs and CRs provides full coverage for the universal quantum gate set. A critical aspect to consider is the larger search space that comes with this approach. The challenge lies in whether we can still identify the desired solutions within this large search space. Experimental results have shown that with a well-designed optimization loop, it is possible to achieve results that surpass those of the gate-level circuits. This suggests that the pulse ansatz can offer sufficient entanglement, expressivity and good coverage of the universal quantum gate set at the same time.}
\rebuttal{In the process, each new step appends a new group of native pulses, which are subsequently updated through a series of iterations with the optimizer. These iterations only update the parameters within the newly appended pulses. At the same time, the pulses established in previous steps are kept in their current state and remain unchanged.}
Such progressive manner resembles the idea of \textbf{quantum speed limit}, which denotes the maximum rate of evolution of a quantum system, i.e., the minimum time required for a quantum system to evolve between two quantum states~\cite{pires2016generalized}. 
Therefore, the first step of \name~might fails to reach a good enough quantum state. 
But it will gradually approximate the desired state with more steps. 
As for the condition to terminate the progressive growth, we propose to stop growing the ansatz when accuracy cannot benefit from the appended step.}

% \paragraph{Pruning}
\noindent\textbf{Pruning.}
We propose to prune the pulse ansatz by removing pulses with parameters that are closest to zeros.
The purpose of pruning is to further reduce the overall pulse latency and decoherence error.
In addition, controls are less robust when their amplitudes are close to zeros.
It is possible that pruning might force \name~to search in a sub-optimal space, but the progressive way can compensate for possible accuracy degradation.
\red{Experimental results show that pruning can significantly reduce the overall latency by an average of 24.13\% on small molecule VQE tasks such as $H_2$, $HeH+$, and $LiH$, while accuracies experience minimal-to-none decreases.}

% We propose to further remove redundant parameters, i.e., redundant pulses, to achieve shorter duration, lower overhead, and lower noise. 
% For \name, firstly, some pulses containing amplitudes close to zero can be safely removed by trimming and fine-tuning. Second, the suboptimal nature of progressive pulse learning leaves room for further optimization of the search circuit by reducing the number of gates.

% As shown in Figure \ref{progressive}, specifically, we first train the searched pulse circuit from zero, collect all the amplitudes, and then remove the pulses where the amplitudes closest to zero are located. We ensure no degradation in the performance of the noise-free simulation compared to the unpruned pulse circuit. Thus, in NISQ machine experiments, the pruned pulse circuits show a small improvement in accuracy due to the reduction of a small number of noise sources.
\begin{table*}[t]
    \centering
    \renewcommand*{\arraystretch}{1.4} 
    \setlength{\tabcolsep}{15pt}
    \footnotesize
    \begin{tabular}{|c|c|c|c|c|} 
        \hline
        Model & Single Qubit Channel & Two Qubit Channel & Training Framework & Compatible with Real Machine \\ \hline
        \name & Yes & Yes & Progressive & Yes \\ \hline
        Ctrl-VQE\cite{meitei2021gate} & Yes & No & Brute-Force & No \\ \hline
    \end{tabular}
    \caption{``Ctrl-VQE'' only considers single-qubit pulse channel. Two-qubit pulses are missing in the ``Ctrl-VQE'' framework, making it inferior for certain tasks. Brute-force optimization makes ``Ctrl-VQE'' limited to small-size problems.}
    \label{functioncomp}
\end{table*}

\noindent\textbf{\red{Analysis of Scalability of \name.}}
The pulse-level optimization algorithm is not compatible with gradient-based methods like parameter-shift rules at the gate level on real quantum machines. 
That is, we cannot obtain the gradients for pulse parameters from the backends of NISQ machines. 
% Non-gradient optimizer becomes the only choice for us to use in the \name~framework on NISQ machine. 
We use COBYLA in \name, while the selection of non-gradient optimizer remains an open problem.
Non-gradient optimizers have two common constraints: they can only handle a limited number of parameters and they do not guarantee convergence to the global optimum.
Therefore, we propose training the ansatz in a ``progressive'' manner, which helps to limit the number of parameters being trained at the same time. Pulse-level ansatz has more parameters than gate-level ansatz but with a much shorter duration, which means pulse ansatz costs less on quantum resources but more on classical computational resources, the trade-off is worth especially in the NISQ era since the decoherence time is limited. A large number of parameters is the intrinsic problem for VQE algorithms, which is not the focus of this work, but we will think in-depth about some good solutions to this problem.

We summarize existing state of art work compared with \name~in Table \ref{functioncomp}.
``Ctrl-VQE'' \cite{meitei2021gate}, where pulse segments are built and the parameters are trained all together.
When ``Ctrl-VQE'' is proposed, it only considers single-qubit pulse channel.
Information of pulse phases and two-qubit pulses are also missing in the ``Ctrl-VQE'' framework, making it infeasible on real quantum devices.
Considerations of two-qubit pulse channel and adaptivity of ansatz size provide \name~ with essentials advantages over ``Ctrl-VQE''.

\revise{The scalability of parameterized pulses is jointly determined by two factors: the number of parameters and the performance of the optimizer.
Firstly, the number of parameters in parameterized pulses scales linearly with the number of pulses. This scaling is the same as the number of parameters in parameterized gates.
One advantage of parameterized pulses is their capacity to integrate a larger number of parameters within the same circuit latency. 
However, it's not always necessary to utilize all available parameters during the optimization process. The flexibility in the number of parameters means we can control the parameter count in pulses. This adaptability allows for different options: we can opt for a higher degree of parameterization to gain more flexibility or we can choose a more conservative approach, aligning the number of parameters with those in gate-level circuits to merely reduce circuit latency.}

\revise{In gate-level VQE tasks, gradient-based optimizers are available with gradients derived through parameter-shift rules. At the pulse level, the concepts of gradient-based optimizers have recently been introduced\cite{peng2023simuq}. To the best of our knowledge, these gradient-based pulse optimizers currently face to challenges to implement on real quantum devices. We would like to point out that the methods in \name~are orthogonal to the choice of optimizers. \name~can be implemented with gradient-based optimizers since it effectively mitigates the problem of high-dimensional parameter space. In conclusion, techniques in \name~should work for both gradient-based and non-gradient optimizers.}

% In \name, where pulse ansatz are built and trained in a ``progressive'' way. In this manner, only a portion of parameters are trained simultaneously.
 
% The non-gradient optimizer has been proved effective with less than 30 parameters.
% Under such circumstance, \name~provides the design space for each step, which can limited the parameters within the ability threshold of optimizer, and the theory is also fit with the QSL. Ctrl-VQE~\cite{meitei2021gate} has proved the pulse-ansatz obtain the power over gate-ansatz, however, it neither consider two-qubit native pulse or the issue of the ability of non-gradient optimizer.
% \usepackage{graphicx}
% \begin{table*}[t]
% \centering
% \renewcommand*{\arraystretch}{1}
% \setlength{\tabcolsep}{8pt}
% \footnotesize
% \begin{tabular}{|c|c|c|c|c|}
% \hline
% Model                & Single Qubit Channel & Two Qubit Channel & Training Framework & Compatiable with Real Machine \\ \hline
% \name & Yes                  & Yes               & Progressive      & Yes  \\ \hline
% Ctrl-VQE\cite{meitei2021gate}             & Yes                  & No                & Brute-Force      & No  \\ \hline
% \end{tabular}%
% \caption{``Ctrl-VQE'' is only considers single-qubit pulse channel.
% Two-qubit pulses are missing in the ``Ctrl-VQE'' framework, making it inferior for certain tasks.
% Brute-force optimization makes ``Ctrl-VQE'' limited to small-size problems.}
% \label{functioncomp}
% \end{table*}

\vspace{-3mm}
\red{
}

\section{Evaluation}
\label{sec6}
\subsection{Backend configuration}
Our experiments are conducted both on simulators and NISQ machines.
Simulators are used to validate that \name~considers the system models of quantum backends.
Simulations are run on a server with two Intel Xeon E5-2630 CPUs (8 cores/CPU), 64 GB DRAM, with CentOS 7.4 as the operating system.
To confirm that our framework works on NISQ machines, we conduct experiments on six IBM's quantum systems: $ibm\_cairo$, $ibmq\_montreal$, $ibmq\_toronto$, $ibmq\_mumbai$,  $ibmq\_guadalupe$ and $ibmq\_jakarta$. 

\subsection{Compilation Setting}
To illustrate our compilation settings and overhead, we show an example task of $H_2$ in the Table \ref{settings}. In our progressive method, each step contains up to 50 iterations. In each iteration, quantum programs are measured with 1024 shots. 
The complexity of the task determines the number of steps that are needed.
We also give the range of amplitudes and frequencies in the table.
The hyperparameter Rhobeg of the optimizer COBYLA is set to 0.1.
\begin{table}[h]
    \centering
    \renewcommand*{\arraystretch}{1.4} 
    \setlength{\tabcolsep}{25pt} 
    \footnotesize
    \begin{tabular}{|c|c|} 
        \hline
        Compilation Settings & Values \\ \hline
        \#Parameters & 6* \\ \hline
        \#Measurement per iteration & 1024 \\ \hline
        \#Iteration per step & 50 \\ \hline
        \#Steps per task & 2* \\ \hline
        Range of Amplitudes & 0 - 0.4 \\ \hline
        Range of Frequencies & -2MHz - 2MHz \\ \hline
        Rhobeg & 0.1 \\ \hline
    \end{tabular}%
    \caption{Example of compilation settings for VQE task of the $H_2$. * means the parameter is dependent on the task. Progressive learning framework determines the number of steps with respect to task definition. }
    \label{settings}
   \vspace{-3mm}
\end{table}

\subsection{\revise{Experiment Setup}}
We choose VQE problems as our evaluation tasks, which consist of several molecules including $H_2$, $HeH+$, $LiH$, $CO_2$, $H_2O$, and $NaH$\revise{, and a three-regular six-nodes graph Maxcut task. For molecular tasks, the `STO-3G' basis set is employed to approximate the spin-orbitals of molecules, which are then transformed into fermionic terms. These terms undergo the Jordan-Wigner (JW) transformation to yield Pauli strings. VQE is then utilized to estimate the ground state energy based on these Pauli strings. In the case of the Maxcut problem, the problem is initially reformulated as a quadratic program, from which the corresponding Ising Hamiltonian is derived. This allows a VQE approach to solve the Maxcut problem using the deduced Ising Hamiltonian.}  

\revise{For the Maxcut problem, the primary metric of interest is the approximation ratio. It's the value achieved by the VQE solution divided by the value of the objective function for the optimal solution. A higher approximation ratio indicates a closer approximation to the optimal solution, and thus better performance of the algorithm.
In the case of molecule ground state energy estimation, the key metric is accuracy. This is typically assessed by comparing the energy computed by the VQE algorithm with the Full Configuration Interaction (FCI) results. FCI is a highly accurate quantum chemistry method that calculates the exact ground state energy of a quantum system within a given basis set. The closer the estimation result is to the FCI benchmark, the more accurate the VQE algorithm is considered to be for molecule tasks.}

\begin{table*}[t]
    \centering
    \renewcommand*{\arraystretch}{1.4} % 调整行高以保持一致性
    \setlength{\tabcolsep}{7pt} % 调整列间距以保持一致性
    \footnotesize
    \begin{tabular}{|c|c|c|c|c|c|c|c|} % 添加竖线以保持边界一致
        \hline % 添加水平线
        \multicolumn{2}{|c|}{Model} & Cairo & Montreal & Toronto & NISQ machine Avg & Simulator & FCI \\ \hline
        \multicolumn{1}{|c|}{\multirow{3}{*}{$H2$}} & Step I & -1.093 (3.870\%) & -1.087 (4.398\%) & -1.073 (5.629\%) & -1.084 (4.661\%) & -1.121 (1.407\%) & -1.137 \\ \cline{2-8} 
        \multicolumn{1}{|c|}{} & Step II & -1.107 (2.639\%) & -1.110 (2.375\%) & -1.073 (5.629\%) & -1.097 (3.518\%) & -1.123 (1.231\%) & -1.137 \\ \cline{2-8} 
        \multicolumn{1}{|c|}{} & Inaccuracy Reduction & 31.83\% & 46.00\% & 0.000\% & 24.52\% & 12.51\% & - \\ \hline
        \multicolumn{1}{|c|}{\multirow{3}{*}{$HeH+$}} & Step I & -2.813 (1.746\%) & -2.845 (0.663\%) & -2.820 (1.485\%) & -2.826 (1.292\%) & -2.855 (0.279\%) & -2.863 \\ \cline{2-8} 
        \multicolumn{1}{|c|}{} & Step II & -2.833 (1.047\%) & -2.866 (0.105\%) & -2.834 (1.013\%) & -2.844 (0.664\%) & -2.856 (0.244\%) & -2.863 \\ \cline{2-8} 
        \multicolumn{1}{|c|}{} & Inaccuracy Reduction & 40.03\% & 84.16\% & 31.78\% & 48.61\% & 12.54\% & - \\ \hline
    \end{tabular}%
    \caption{Results of Estimated Energy for Molecules in Different Steps.}
    \label{stepcompare}
\end{table*}
\begin{table*}[t]
    \centering
    \renewcommand*{\arraystretch}{1.4} % 调整行高以保持一致性
    \setlength{\tabcolsep}{7pt} % 调整列间距以保持一致性
    \footnotesize
    \begin{tabular}{|c|c|c|c|c|c|c|c|c|} % 添加竖线以保持边界一致
        \hline % 添加水平线
        \textbf{Model} & \textbf{Ansatz Level} & \textbf{Qubits} & \textbf{Duration} & \textbf{SNP Count} & \textbf{CR Count} & \textbf{Molecule} & \textbf{Energy} & \textbf{Reference Energy} \\
        \hline
        Random Generated Ansatz & Gate Ansatz & 2 & 682.7ns & 16 & 2 & $H_2$ & -0.853 & -1.137 \\
        \hline
        RealAmplitude Ansatz~\cite{qiskit} & Gate Ansatz & 2 & 376.9ns & 12 & 1 & $H_2$ & -0.974 & -1.137 \\
        \hline
        QuantumNAS~\cite{wang2022quantumnas} & Gate Ansatz & 2 & 682.7ns & 16 & 2 & $H_2$ & -1.033 & -1.137 \\
        \hline
        \name~ & Pulse Ansatz & 2 & 71.1ns & 3 & 0 & $H_2$ & -1.100 & -1.137 \\
        \hline
        RealAmplitude Ansatz & Gate Ansatz & 2 & 753.8ns & 24 & 2 & $HeH+$ & -2.691 & -2.863 \\
        \hline
        \name~ & Pulse Ansatz & 2 & 199.1ns & 1 & 1 & $HeH+$ & -2.866 & -2.863 \\
        \hline
        QuantumNAS & Gate Ansatz & 6 & 7296.0ns & 40 & 12 & $LiH$ & -6.914 & -7.882 \\
        \hline
        \name~ & Pulse Ansatz & 4 & 199.1ns & 4 & 2 & $LiH$ & -7.590 & -7.882 \\
        \hline
    \end{tabular}
        \caption{Comparison of duration, pulse counts, and estimated energy of gate ansatz and the native-pulse ansatz generated by \name~on NISQ machines. \rebuttal{SNP} indicates Single-Qubit Native Pulse, and CR indicates Cross-Resonance Pulse.}
    \label{CRcompare}
\end{table*}
\subsection{Native pulse}
In the Table \ref{compare}, we compare the results of different ansatz types for VQE tasks.
The results are collected from NISQ machines.
To determine the capabilities of our native-pulse model, we employ the simplest native-pulse model and compare its performance to that of native gates.
The Table \ref{compare} demonstrates that our native-pulse approach can deliver superior VQE results in less durations.
Pulse ansatz with only two native pulses can produce -1.036H energy, whereas the two-gate ansatz can only achieve -0.534H energy.
The two-gate ansatz consists of both single-qubit and double-qubit gates.
The duration of the pulse ansatz is approximately 30\% less than that of a gate ansatz.
The duration numbers are derived from the pulses that are returned by quantum backends.
In conclusion, we are able to produce superior VQE results with significantly less latency on NISQ machines.

\subsection{Simulation results}
To confirm that our framework can generate accurate energy curves for given molecules, we use simulators to evaluate our methodologies.
The specifications of ``fake'' backends are collected by the simulators and used as Hamiltonian configurations to run the pulses. 
\red{
We use pulse simulator that is provided by IBM qiskit toolkit. It simulates continuous time Hamiltonian dynamics of a quantum system, with controls specified by pulses.
}
\begin{figure}[t]
\centering
\includegraphics[width=0.97\linewidth]{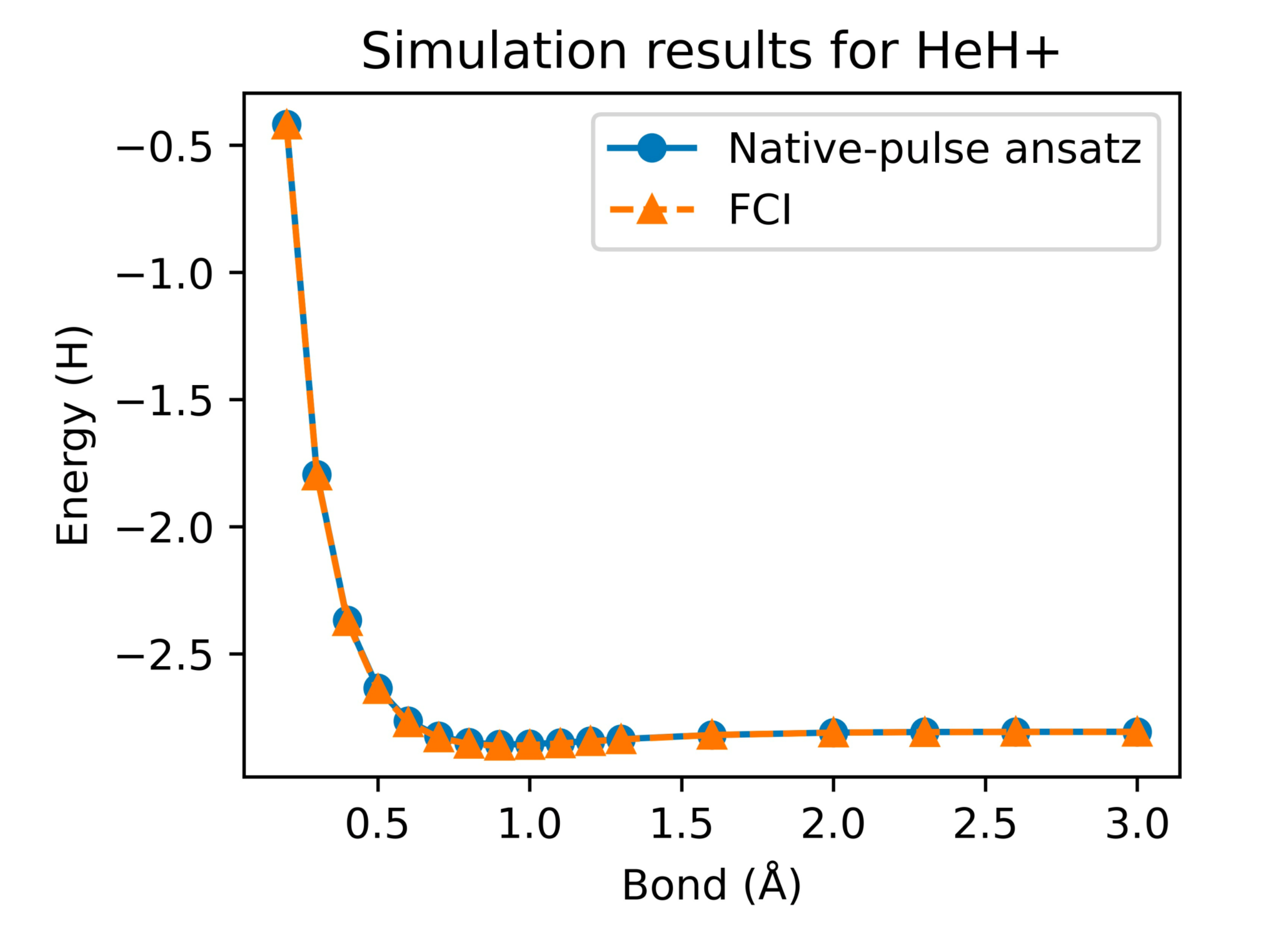}
% \vspace{-1mm}

\caption{Simulation results of $HeH+$ molecule. The pulse ansatz curve almost matches the FCI value. The lowest energy point reaches an accuracy of 99.756\%.}
\label{simulationHeH}
\vspace{-3mm}
\end{figure}
The simulation results in Figure \ref{simulationHeH} indicate that our methods can produce comparable results to those of the FCI methods.
When attempting to compute the ground state energy of HeH+ molecules, we discover that the simulated results are only 0.244\% off from the FCI results.
In the case of H2 molecules, our simulated results deviate from FCI data by 1.23\%.
As shown in Table \ref{stepcompare}, the insertion of a native pulse in STEP two improved the performance of the native-pulse ansatz by approximately 12\% on simulators. 
Overall, simulation results validates the methodology of \name~.

\begin{figure*}[t]
\centering
\includegraphics[width=0.98\linewidth]{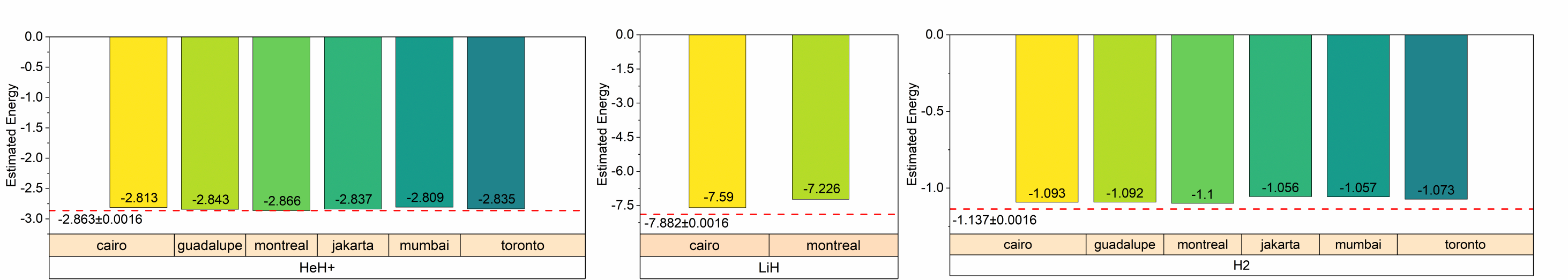}
\caption{Evaluation of native-pulse ansatz on NISQ machines for $H_2$, $HeH+$, and $LiH$ VQE tasks. NISQ machines include $ibm\_cairo$, $ibmq\_montreal$, $ibmq\_toronto$, $ibmq\_mumbai$,  $ibmq\_guadalupe$, and $ibmq\_jakarta$. Our toy ansatz can achieve great accuracy on all the NISQ machines. For $ibmq\_montreal$ where we obtain the best results, the average \rebuttal{CX} error is 1.518\% as the average readout error is 3.457\%. \name~is proven to be robust and error-resilient on NISQ machines.}
%\vspace{-3mm}
\label{evaluation}
\end{figure*}

\begin{figure*}[htb!]
\centering
\includegraphics[width=\linewidth]{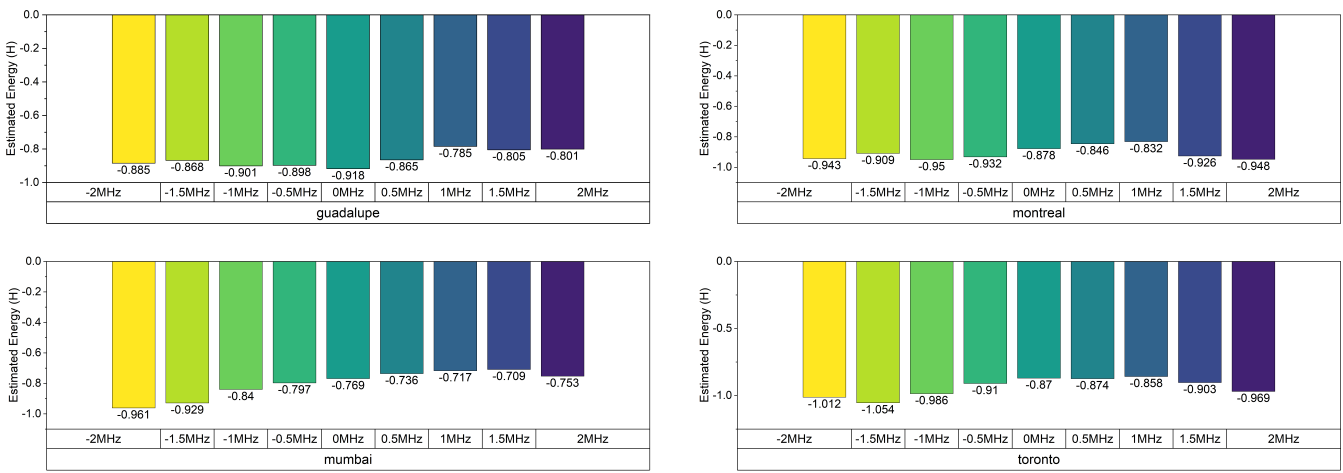}
\caption{Influence of frequency shift on the estimated energies. 
Even with a narrow frequency shift range, we notice obvious changes in the estimated energies. 
% The experiments are conducted before convergence, the energy values are calculated for $H_2$ molecule.
The experimental results show that \revise{for the $H_2$ molecule's ground state energy estimation task,} as we vary the detuning frequency, the energy estimation scans over a broad range, which validates our methods to tune frequency in pulses.
}
\label{shfitfreq}
\end{figure*}

% \begin{table*}[t]
% \centering
% \renewcommand*{\arraystretch}{1}
% \setlength{\tabcolsep}{6pt}
% \footnotesize
% \caption{Results of Estimated Energy for Molecules in Different Steps}
% \begin{tabular}{cc|c|c|c|c|c|c}
% \midrule
% \multicolumn{2}{c|}{Model}                   & Cairo            & Montreal         & Toronto          & NISQ machine Avg & Simulator        & FCI    \\ \midrule
% \multicolumn{1}{c|}{\multirow{3}{*}{$H2$}}   & Step I & -1.093 (3.870\%) & -1.087 (4.398\%) & -1.073 (5.629\%) & -1.084 (4.661\%) & -1.121 (1.407\%) & -1.137 \\ \cmidrule{2-8} 
% \multicolumn{1}{c|}{} & Step II              & -1.107 (2.639\%) & -1.110 (2.375\%) & -1.073 (5.629\%) & -1.097 (3.518\%) & -1.123 (1.231\%) & -1.137 \\ \cmidrule{2-8} 
% \multicolumn{1}{c|}{} & Inaccuracy Reduction & 31.83\%          & 46.00\%          & 0.000\%          & 24.52\%          & 12.51\%          & -      \\ \midrule
% \multicolumn{1}{c|}{\multirow{3}{*}{$HeH+$}} & Step I & -2.813 (1.746\%) & -2.845 (0.663\%) & -2.820 (1.485\%) & -2.826 (1.292\%) & -2.855 (0.279\%) & -2.863 \\ \cmidrule{2-8} 
% \multicolumn{1}{c|}{} & Step II              & -2.833 (1.047\%) & -2.866 (0.105\%) & -2.834 (1.013\%) & -2.844 (0.664\%) & -2.856 (0.244\%) & -2.863 \\ \cmidrule{2-8} 
% \multicolumn{1}{c|}{} & Inaccuracy Reduction & 40.03\%          & 84.16\%          & 31.78\%          & 48.61\%          & 12.54\%          & -      \\ \midrule
% \end{tabular}%
% \label{stepcompare}
% \end{table*}

\subsection{NISQ machine results}
Our techniques are also evaluated on several NISQ computers.
As presented in Figure \ref{evaluation} and Table \ref{CRcompare}, a toy model of native-pulse ansatz is capable of producing promising results. In the instance of the HeH+ molecule, we attain an average accuracy of 99.336\%, with 99.895\% being the highest achievable accuracy.
The absolute difference in energy is 0.003H. 
The accuracy is close to the requirement of computational chemistry (0.0016H), which is the chemical accuracy constant for computational chemistry. 
It qualifies the typical minimum energy gap that can be verified through experiments.
As for the H2 molecule, we attain an average accuracy of 96.482\% and a maximum accuracy of 97.625\% with native-pulse ansatz.
For LiH, a bigger molecule, we obtain an energy accuracy of 96.295\%.
These accuracy figures are derived from NISQ machines with gate and measurement errors that exceed 1\%.
We observe that $ibmq\_montreal$ tends to return the best results, while $ibmq\_mumbai$ typically returns the worse results.
Then, we confirm that $ibmq\_mumbai$ has larger error rates for gates and measurements on average than $ibmq\_montreal$.

The results collected from NISQ computers demonstrate that our approaches are highly error-tolerant.
In terms of total duration time, our native-pulse approaches have significant advantages over the current gate ansatz generator.
To ensure equality, the gate ansatz of the baselines is implemented on the same NISQ computer, and their duration is determined with the acquired pulse schedules.
We are able to demonstrate a 97.3\% reduction in ansatz duration compared to QuantumNAS, and our estimated energy numbers for LiH are lower.
With a duration decrease of 89.6\%, our method can obtain better energy values while dealing with H2 molecules.
QuantumNAS does not report the energy numbers for the HeH+ molecule. Thus we compare our techniques to the Real Amplitude Ansatz, which shows that we are able to reduce duration by 73.6\% while maintaining similar performance. 
\begin{figure}[t]
\centering
\includegraphics[width=0.97\linewidth]{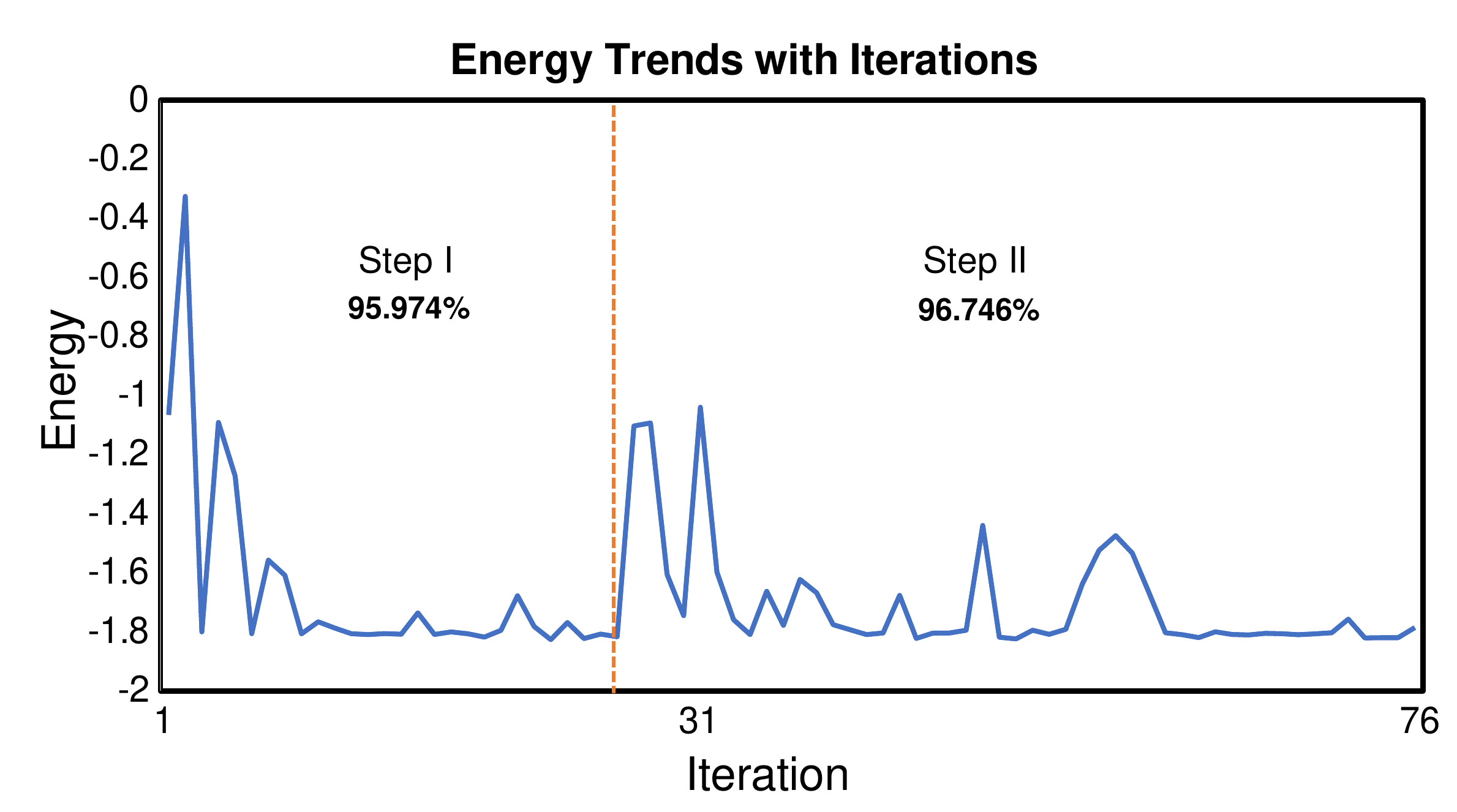}
\caption{Energy trends with \#iterations for a $H_2$ VQE task. The data is collected from $ibmq\_montreal$. \name~with non-gradient optimizer for pulse-level optimization. The vertical dotted line in the middle separates the steps in progressive learning. We can see several ``peaks'' on the curves because non-gradient optimizer attempts might be made in a bad ``direction''. \revise{The accuracy achieved in Step I, encompassing optimization iterations one through 26, is 95.974\%. Subsequently, Step II, spanning iterations 27 to 76, further enhances the accuracy to 96.746\%.}}
\label{trendscompare}
\end{figure}
\begin{figure}[t]
\centering
\includegraphics[width=0.97\linewidth]{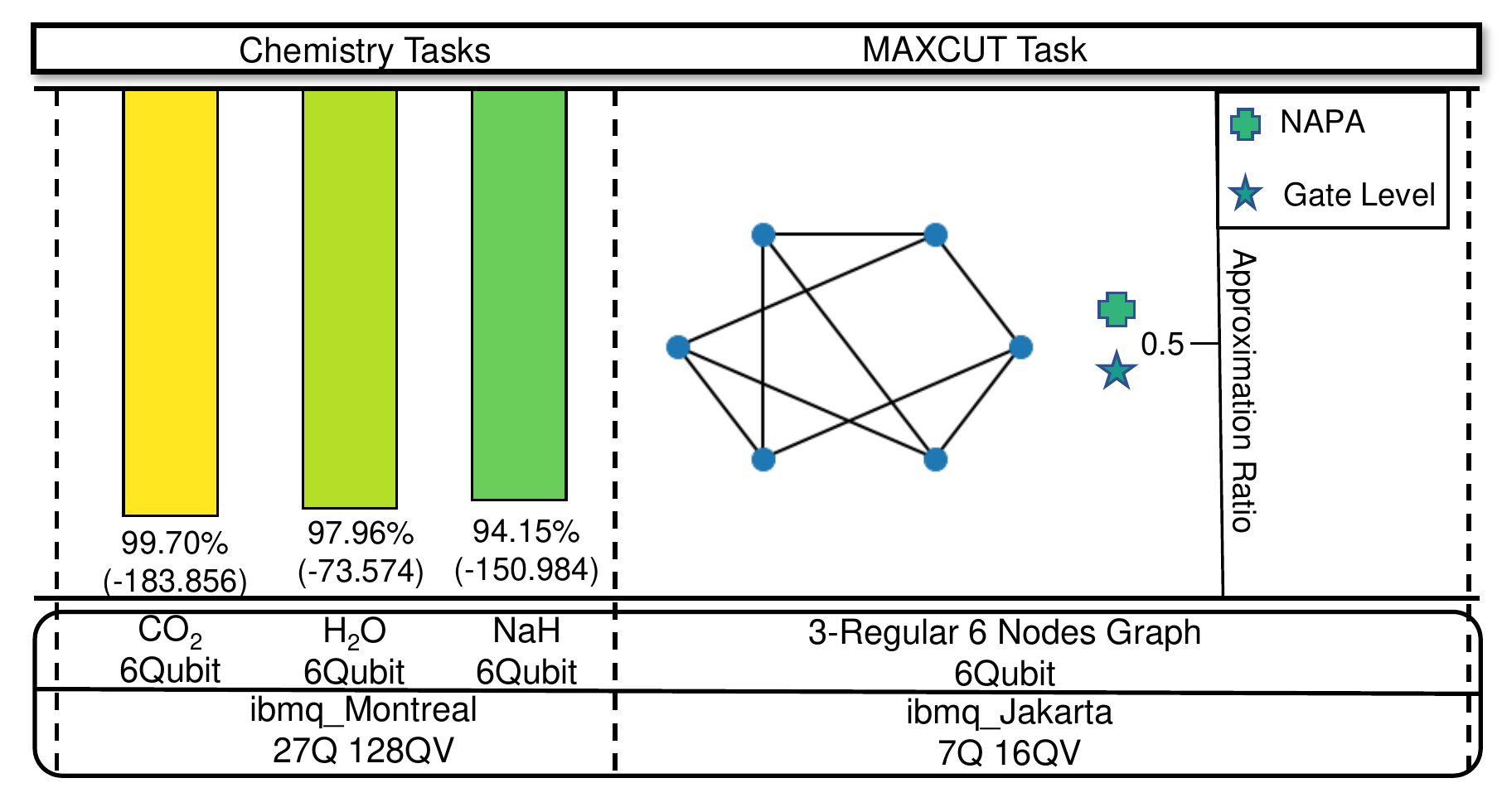}
\caption{\red{Evaluation of native-pulse ansatz on NISQ machines for $CO_2$, $H_2O$, and $NaH$'s VQE tasks as well as three-regular six-nodes graph maxcut task. NISQ machines include $ibmq\_montreal$ and $ibmq\_jakarta$. \revise{``Gate level'' in the Maxcut task indicates the ``TwoLocal'' ansatz from Qiskit circuit library which consists of Ry and CZ gates.}}}
\label{additionalevaluation}
\end{figure}

% \subsection{Tuning frequencies}
% RENHANG TODO
\rebuttal{In Fig. \ref{shfitfreq}}, we detune the pulse frequency and use a few steps to estimate the energy of the H2 molecule. The experimental results show that as we vary the detuning, the energy estimation scans over a broad range, which ensures that the pulse parameter variation in the learning step is wide enough to cover the ground energy. Moreover, the evident response of energy estimation with pulse parameterization shows that our ansatz structures are valid in covering all spin-orbital configurations of the electronic structure. The estimated energy can then progressively approximate the desired ground energy through pulse parameter learning.

% \subsection{}
Figure \ref{trendscompare} depicts the relationships between the computed energy and the number of iterations.
Despite the fact that it is not visible in Figure \ref{trendscompare}, we validate that our progressive approaches work on VQE tasks.
Since a non-gradient optimizer is deployed, we can observe several peaks of the curve where the non-gradient optimizer attempts to update the pulse ansatz's parameters but obtains poorer results.
As indicated in the Table \ref{stepcompare}, we are able to reduce the deviation from FCI values by around 40\%, when native-pulses ``grow'' progressively.
Only in one instance where the pulse ansatz were conducted on $ibmq\_toronto$ were the results not improved.
Taking into account this failure, we find a 24\% improvement for H2 molecule experiments and a 48\% improvement for HeH+ molecule experiments. 
\red{
We also evaluate \name~on larger chemistry tasks as well as optimization tasks as shown in Fig. \ref{additionalevaluation}. For $CO_2$, $H_2O$, and $NaH$, the pulse ansatz generated by \name~has a duration of 1031.1 ns on $ibmq\_montreal$ and the estimated ground state energies are -183.856H (99.70\% accuracy), -73.574H (97.96\% accuracy), and -150.984H (94.15\% accuracy), respectively. For a three-regular six-node graph, the experiment is conducted on NISQ machine $ibmq\_jakarta$, \revise{\name~produces the pulse ansatz} with resulting improved approximation \rebuttal{ratio} of 3.4\% compared to a gate-level ansatz of similar structure. \revise{We use the `TwoLocal' ansatz which is a type of HEA that consists of Ry and CZ gates as our gate-level baseline.}. Results demonstrate the feasibility of \name~on different \rebuttal{real-world applications}. 
}

% \jlc{for a ansatz with sub-optimal }

\section{Conclusion}
\label{sec8}
\name~is a framework for constructing native-pulse ansatz for VQAs.
As a result of removing an abstraction layer of native gates, native pulses provide huge latency advantages.
Then, we employ progressive learning to ``grow'' our ansatz. 
Thus, our pulse ansatz is better able to explore the Hilbert space, while the optimizer is still able to handle the problem.
Extensive experiments are conducted on NISQ machines, and the results indicate that an average of 86\% reduction in latency is obtained with up to 99.895\% accuracy on small molecule VQE tasks.
\red{Experiments on larger-size molecule VQE tasks achieve an average accuracy of 97.27\%.
The results of \name~on maxcut problem demonstrate the feasibility of \name~on different problems. 
How to explore the applications of \name~on different types of VQAs remains an open question.
Also, potentials of \name~on quantum algorithms other than VQAs are to be investigated.
Overall, our experimental results show that pulse-level methods like \name~can greatly reduce the duration of quantum circuits, thus enhance the capabilities of quantum computers.
% Methods like \name~can help enhance the capabilities of quantum machines.
% And we believe that in the future, we will be able to investigate the strength of quantum algorithms other than VQA on the NISQ machine. 
%For instance, the ability of Shor's algorithm is constrained by the decoherence time of the current NISQ machine. NAPA has the ability to circumvent this drawback, may allowing us to solve large problems with Shor's algorithm on NISQ machines.
}
\section{Acknowledgments}
This work is funded in part by EPiQC, an NSF Expedition in Computing, under award CCF-1730449; in part by STAQ under award NSF Phy-1818914/232580; in part by the US Department of Energy Office of Advanced Scientific Computing Research, Accelerated 
Research for Quantum Computing Program; and in part by the NSF Quantum Leap Challenge Institute for Hybrid Quantum Architectures and Networks (NSF Award 2016136), in part based upon work supported by the U.S. Department of Energy, Office of Science, National Quantum 
Information Science Research Centers, and in part by the Army Research Office under Grant Number W911NF-23-1-0077. The views and conclusions contained in this document are those of the authors and should not be interpreted as representing the official policies, either expressed or implied, of the U.S. Government. The U.S. Government is authorized to reproduce and distribute reprints for Government purposes notwithstanding any copyright notation herein.

FTC is the Chief Scientist for Quantum Software at Infleqtion and an advisor to Quantum Circuits, Inc.

%%%%%%% -- PAPER CONTENT ENDS -- %%%%%%%%

%%%%%%%%% -- BIB STYLE AND FILE -- %%%%%%%%
\bibliographystyle{IEEEtranS}
\bibliography{refs.bib}
\vspace{-10mm}
%%%%%%%%%%%%%%%%%%%%%%%%%%%%%%%%%%%%
\begin{IEEEbiography}[{\includegraphics[width=1in,height=1.25in,clip,keepaspectratio]{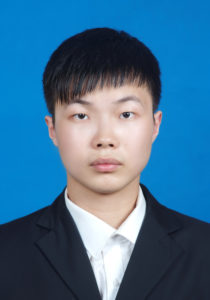}}]{Zhiding Liang}
is a third-year Ph.D. student at the University of Notre Dame, Department of Computer Science and Engineering, advised by Prof. Yiyu Shi. His research focuses on quantum pulse control, hardware/software codesign for quantum computing, and VQAs. His research has been recognized as the DAC Young Fellow twice, both in 2020 and 2021, and highlighted by IBM Qiskit, and appears in top conferences such as DAC, ICCAD, and QCE. He is the recipient of the Edison Innovation Fellowship. 
\end{IEEEbiography}
\vspace{-13mm}
\begin{IEEEbiography}[{\includegraphics[width=1in,height=1.25in,clip,keepaspectratio]{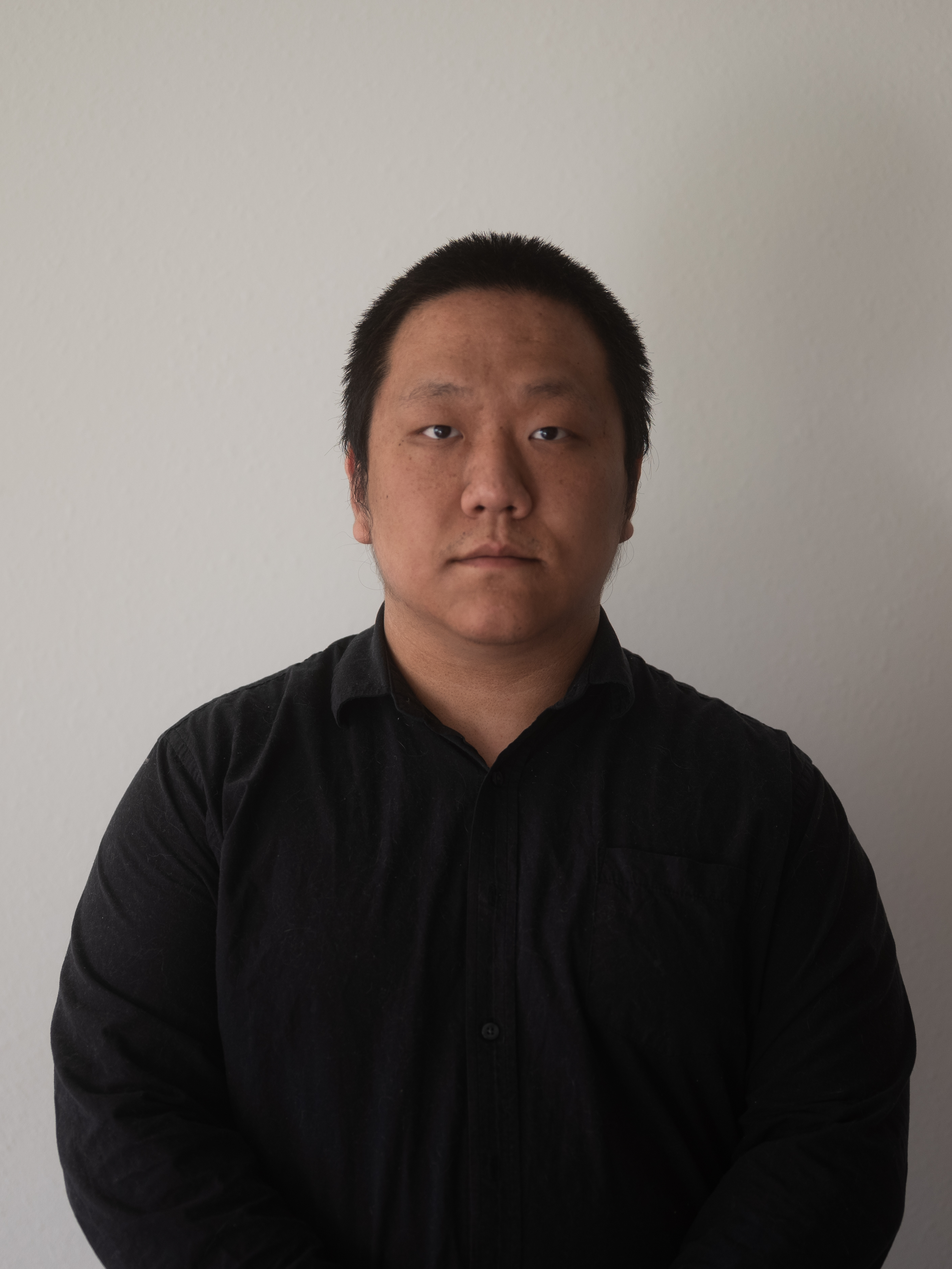}}]{Jinglei Cheng}
is a final-year Ph.D. student at Purdue University, Department of Computer Science. He is advised by Prof Xuehai Qian in the ALCHEM Group. He received Bachelor of Engineering Degree in the Department of Microelectronic Science and Engineering at Tsinghua University in 2018. His research focuses on VQAs and quantum computer architecture. And his research papers appear in conferences including MICRO, ISCA, DAC and QCE.
\end{IEEEbiography}
\vspace{-13mm}
\begin{IEEEbiography}[{\includegraphics[width=1in,height=1.25in,clip,keepaspectratio]{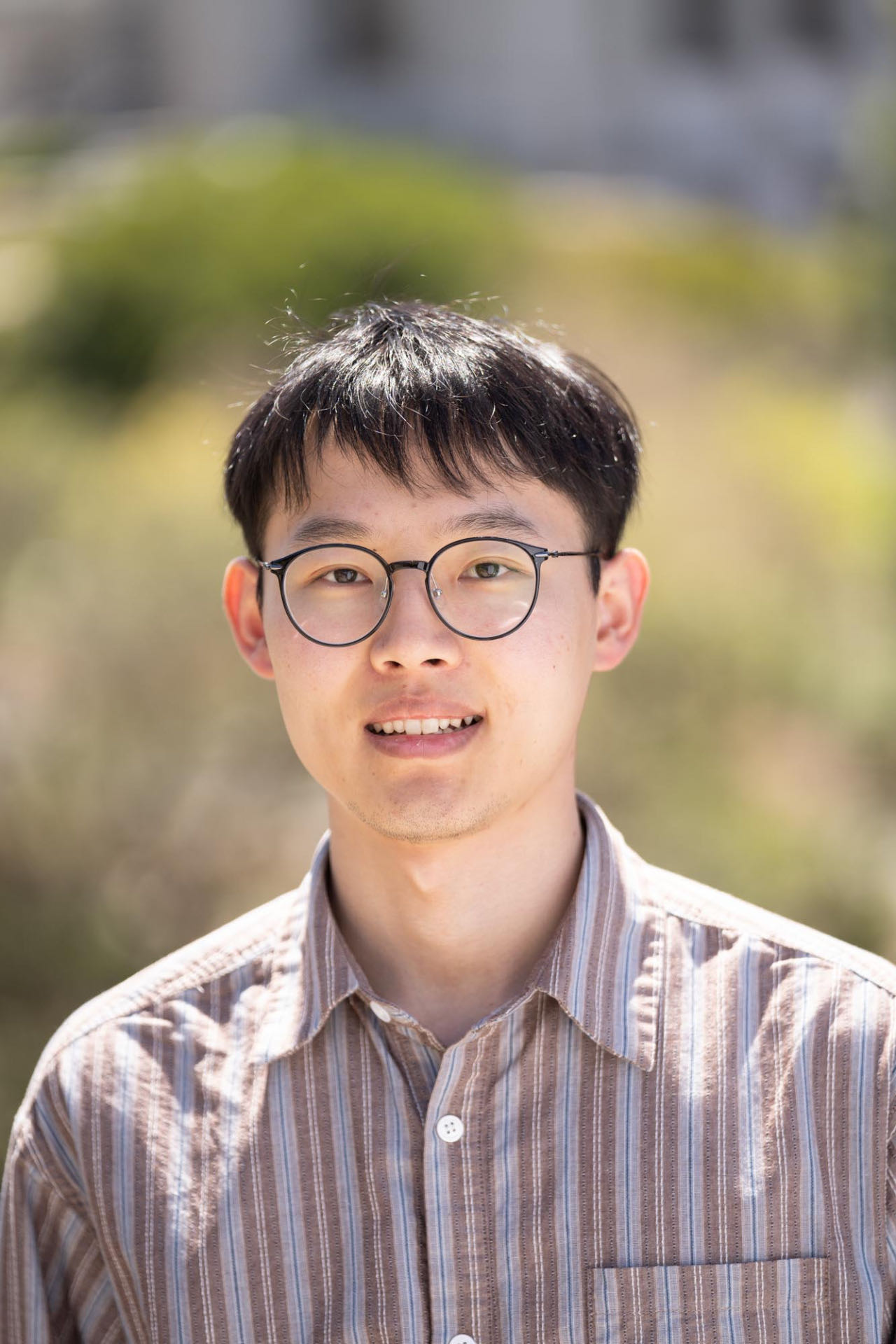}}]{Hang Ren}
is currently a Ph.D. student at the University of California, Berkeley. He received Bachelor's degree from Nankai University. His research is about quantum error mitigation and shadow tomography. 
\end{IEEEbiography}
\vspace{-13mm}
\begin{IEEEbiography}[{\includegraphics[width=1in,height=1.25in,clip,keepaspectratio]{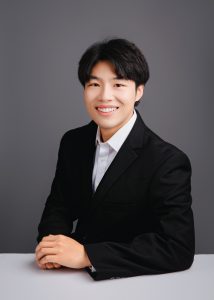}}]{Hanrui Wang}
Hanrui Wang is a final-year Ph.D. student at MIT EECS advised by Prof. Song Han. His research focuses on quantum computer architecture, ML for quantum and efficient ML. His research has been recognized by ACM student research competition 1st place award, best poster award at NSF AI Institute and appears in top conferences such as MICRO, HPCA, DAC, ICCAD and NeurIPS. He is the recipient of Qualcomm Fellowship, Unitary Fund, and Nvidia Fellowship Finalist. 
\end{IEEEbiography}
\vspace{-13mm}
\begin{IEEEbiography}[{\includegraphics[width=1in,height=1.25in,clip,keepaspectratio]{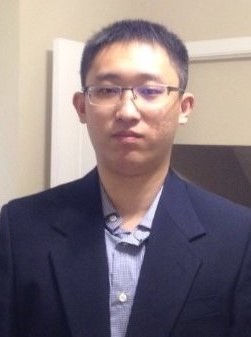}}]{Fei Hua}
is a 6th-year Ph.D. at the Computer Science Department at Rutgers. His major interests are quantum program compilation, quantum error mitigation and correction. Fei got his M.S. degree in Computer Science from Washington University at St. Louis. He did his internships at IBM Research in 2022, and at Pacific Northwest National Lab (PNNL) in 2023. He is one of the lead developers for the open-source compiler QASM Transpiler.
\end{IEEEbiography}
\vspace{-13mm}
\begin{IEEEbiography}[{\includegraphics[width=1in,height=1.25in,clip,keepaspectratio]{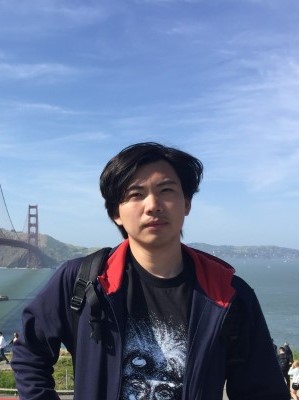}}]{Zhixin Song}
is currently a Ph.D. student at the Georgia Institute of Technology. He received Bachelor's degree from the Ohio State University in 2019. His research interests include near-term quantum algorithms for solving differential equations and computational fluid dynamics (CFD), cooperating HPC systems and quantum hardware.
\end{IEEEbiography}
\vspace{-13mm}

\begin{IEEEbiography}[{\includegraphics[width=1in,height=1.25in,clip,keepaspectratio]{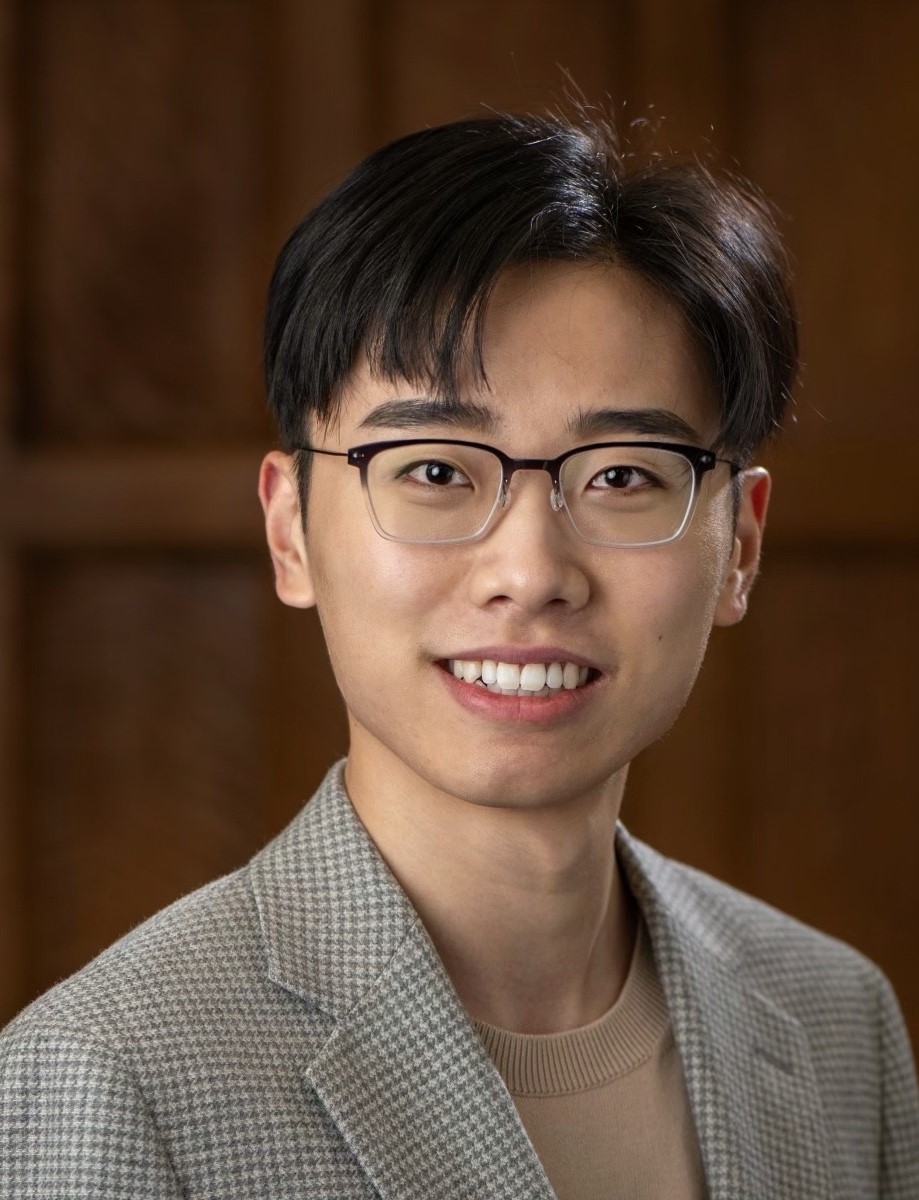}}]{Yongshan Ding}
is an Assistant Professor of Computer Science at Yale University. Ding completed his Ph.D. from the University of Chicago. He is a recipient of the William Rainey Harper Dissertation Fellowship, one of UChicago's highest honors, and the Siebel Scholarship. Before that, he received his B.Sc. degrees in Computer Science and Physics from Carnegie Mellon University. 
\end{IEEEbiography}
\vspace{-13mm}
\begin{IEEEbiography}[{\includegraphics[width=1in,height=1.25in,clip,keepaspectratio]{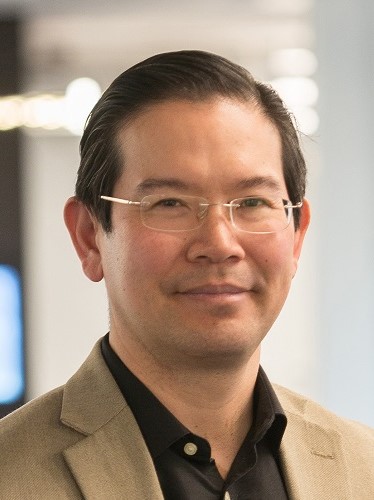}}]{Frederic T. Chong (IEEE Fellow)}
is the Seymour Goodman Professor in the Department of Computer Science at the University of Chicago. He is also Lead Principal Investigator for the EPiQC Project, an NSF Expedition in Computing. Chong received his Ph.D. from MIT in 1996 and was a faculty member and Chancellor's fellow at UC Davis from 1997-2005. He was a Professor of Computer Science, Director of Computer Engineering, and Director of the Greenscale Center for Energy-Efficient Computing at UCSB from 2005-2015. 
\end{IEEEbiography}
\vspace{-13mm}

\begin{IEEEbiography}[{\includegraphics[width=1in,height=1.25in,clip,keepaspectratio]{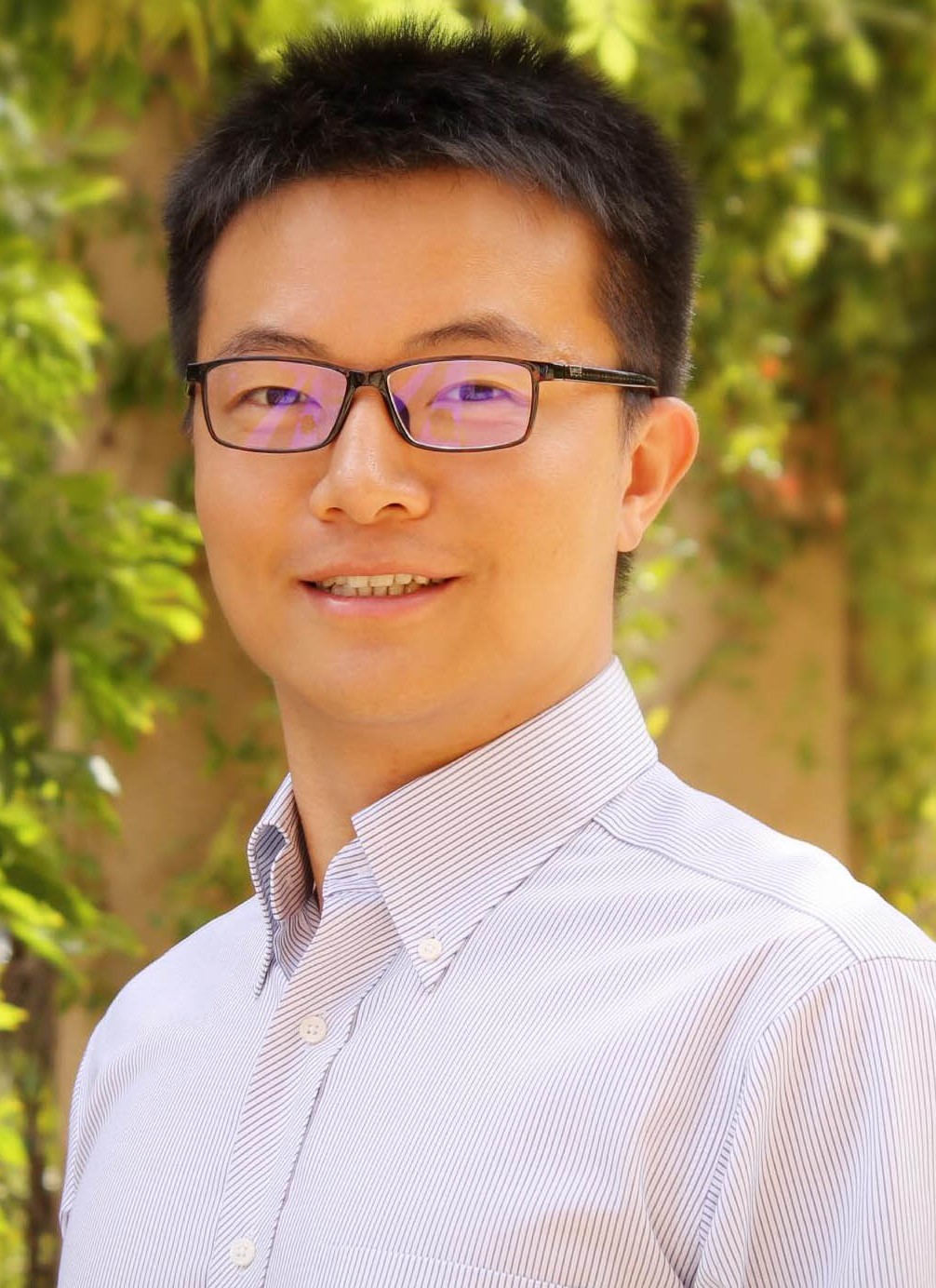}}]{Song Han}
is an associate professor at MIT EECS. He received his PhD degree from Stanford University. He proposed the “Deep Compression” technique including pruning and quantization that is widely used for efficient AI computing, and “Efficient Inference Engine” that first brought weight sparsity to modern AI chips, which influenced NVIDIA’s Ampere GPU Architecture with Sparse Tensor Core. He pioneered the TinyML research that brings deep learning to IoT devices. 
\end{IEEEbiography}
\vspace{-13mm}
\begin{IEEEbiography}[{\includegraphics[width=1in,height=1.25in,clip,keepaspectratio]{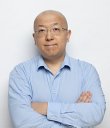}}]{Xuehai Qian}
is an associate professor of the Department
of Computer Science at Purdue University. His research interests include domain-specific systems and architectures, performance tuning and resource management of cloud systems, and parallel computer architectures. He got his Ph.D from the University of Illinois Urbana Champaign. He is the recipient of W.J Poppelbaum Memorial Award at UIUC, NSF CRII and CAREER Award, and the inaugural ACSIC Rising Star Award.
\end{IEEEbiography}
\vspace{-13mm}
\begin{IEEEbiography}[{\includegraphics[width=1in,height=1.25in,clip,keepaspectratio]{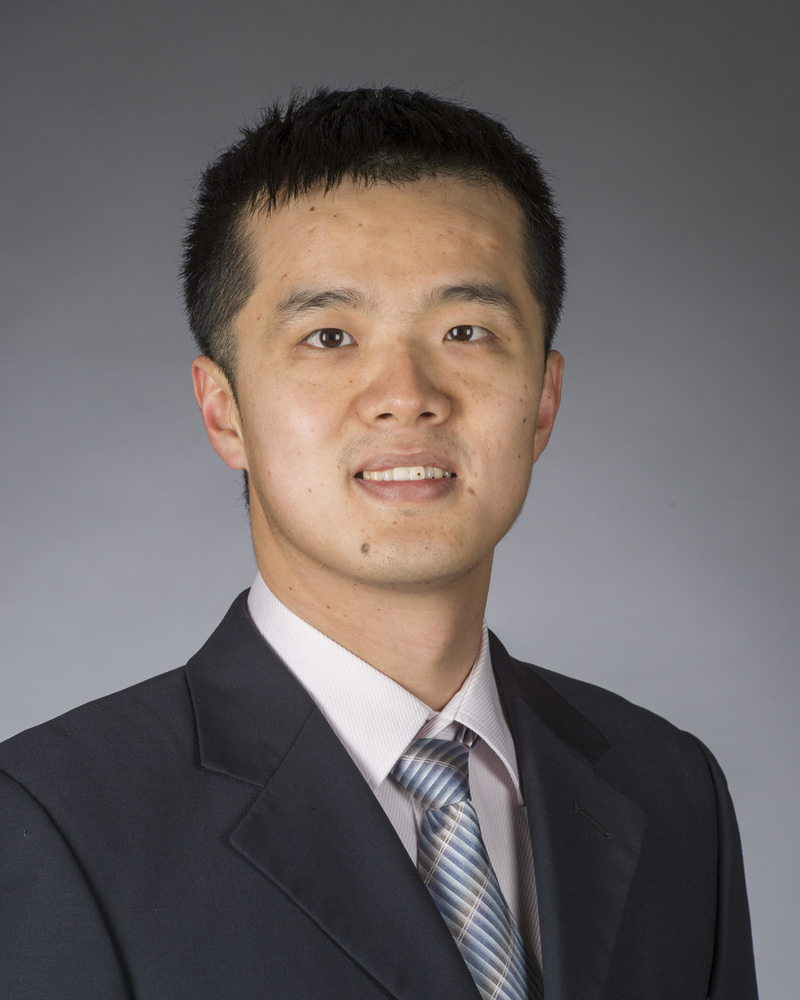}}]{Yiyu Shi}
is currently a professor in the Department of Computer Science and Engineering at the University of Notre Dame, the site director of National Science Foundation I/UCRC Alternative and Sustainable Intelligent Computing, and the director of the Sustainable Computing Lab (SCL). He received his B.S. in Electronic Engineering from Tsinghua University, Beijing, China in 2005, the M.S and Ph.D. degree in Electrical Engineering from the University of California, Los Angeles in 2007 and 2009 respectively. His current research interests focus on hardware intelligence and biomedical applications. 
\end{IEEEbiography}
\end{document}